\documentclass[final, 5p, times, twocolumn, number, sort&compress]{elsarticle}

\usepackage[colorlinks=true,allcolors=blue]{hyperref}
\usepackage[scaled=0.85]{roboto-mono}
\usepackage[T1]{fontenc}
\usepackage{listings}
\usepackage{textcomp}
\usepackage{xcolor}
\usepackage{xparse}
\usepackage{graphicx}
\usepackage{caption}
\usepackage{upgreek}
\usepackage{dcolumn}
\usepackage{amsmath}
\usepackage{amsfonts}
\usepackage{amssymb}
\usepackage{mathtools}
\usepackage{bm}
\usepackage{cuted}
\usepackage{braket}
\usepackage{orcidlink}
\usepackage{nicefrac}
\usepackage{algorithm}
\usepackage{algorithmic}
\usepackage{booktabs}
\usepackage[online,referable]{threeparttablex}
\usepackage{tikz}
\usetikzlibrary{shapes.geometric, arrows.meta, positioning, shadows}
\usepackage{siunitx}
\usepackage{tabularx}

\usepackage{pythonhighlightcustom}

\newcommand{\dfmk}{\texttt{DeepFMKit}}

\begin{document}

\journal{Computer Physics Communications}

\title{The \dfmk{} Python package: A toolbox for simulating and analyzing\\deep frequency modulation interferometers}

\author[1]{M.\ Dovale-\'Alvarez\orcidlink{0000-0002-6300-5226}\corref{cor}}\ead{mdovale@arizona.edu}

\cortext[cor]{Corresponding author}

\affiliation[1]{
  organization={James C.\ Wyant College of Optical Sciences, University of Arizona},
  postcode={85721},
  city={Tucson},
  country={United States}}

\begin{abstract}
Deep Frequency Modulation Interferometry (DFMI) is an emerging laser interferometry technique for high-precision metrology, offering picometer-level displacement measurements and the potential for absolute length determination with sub-wavelength accuracy. However, the design and optimization of DFMI systems involve a complex interplay between interferometer physics, laser technology, multiple noise sources, and the choice of data processing algorithm. To address this, we present \dfmk{}, a new open-source Python library for the end-to-end simulation and analysis of DFMI systems. The framework features a high-fidelity physics engine that rigorously models key physical effects such as time-of-flight delays in dynamic interferometers, arbitrary laser modulation waveforms, and colored noise from user-defined $1/f^\alpha$ spectral densities. This engine is coupled with a suite of interchangeable parameter estimation algorithms, including a highly-optimized, parallelized frequency-domain Non-linear Least Squares (NLS) for high-throughput offline analysis, and multiple time-domain Extended Kalman Filter (EKF) implementations for real-time state tracking, featuring both random walk and integrated random walk (constant velocity) process models. Furthermore, \dfmk{} includes a high-throughput experimentation framework for automating large-scale parameter sweeps and Monte Carlo analyses, enabling systematic characterization of system performance. \dfmk{}'s modular, object-oriented architecture facilitates the rapid configuration of virtual experiments, providing a powerful computational tool for researchers to prototype designs, investigate systematic errors, and accelerate the development of precision interferometry.
\end{abstract}

\begin{keyword}
Deep Frequency Modulation Interferometry \sep 
Laser Interferometry \sep 
Physics Simulation \sep
Parameter Estimation \sep
Scientific Software \sep
Python \sep
Nonlinear Least Squares \sep 
Extended Kalman Filter
\end{keyword}

\maketitle

\section{Introduction}
\label{sec:introduction}

Laser interferometry is a cornerstone of modern science and technology, enabling measurements of unparalleled precision. A prominent application is its use as the core technology behind gravitational wave observatories, such as LIGO~\cite{LIGO,GW150914}, Virgo~\cite{AdvancedVirgo15}, KAGRA~\cite{Akutsu2019}, and the forthcoming Einstein Telescope~\cite{ETDesignReport2020} and LISA mission~\cite{LISA,LISARedBook}. Beyond fundamental science, interferometry is indispensable across diverse technological sectors~\cite{deGroot2019, Huang2025}. Among the various interferometric techniques, Deep Frequency Modulation Interferometry (DFMI) has emerged as an important method for developing compact, high-precision sensing devices~\cite{Gerberding2015, Isleif2016, Isleif2019, Eckhardt2024}. DFMI utilizes the strong, sinusoidal frequency modulation of a laser source injected into an unequal-arm interferometer. The resulting interferometric signal encodes path length information in a rich harmonic structure, offering highly sensitive displacement and absolute length measurements, as well as significant potential for the miniaturization of the optical setup.

The design, characterization, and optimization of a DFMI system, however, present significant computational challenges. The measured signal exhibits a complex, non-linear dependence on numerous system parameters. Its behavior is affected by a multitude of factors, including the laser's intrinsic frequency and amplitude noise, the dynamic motion of the target, non-ideal electronic responses, and distortions in the modulation waveform~\cite{Gerberding2021, Eckhardt2022, An2024, Dovale2025}. Furthermore, system performance is critically dependent on the data analysis algorithm used to extract physical parameters from the raw signal. While various readout strategies exist, from batch-processing Non-linear Least Squares (NLS) fits to real-time Extended Kalman Filters (EKF), a systematic comparison of their performance under different noise conditions and systematic effects requires a flexible and accurate simulation environment. The absence of a unified, open-source framework that integrates high-fidelity physical modeling with a suite of analysis tools has hindered the broader exploration and adoption of this technology.

To fill this gap, we introduce \dfmk{}, an open-source Python package for the end-to-end simulation and analysis of DFMI systems. \dfmk{} provides a comprehensive computational framework enabling researchers to model the entire chain of a DFMI experiment. Its novelty lies in an integrated design that combines a high-fidelity physics engine with a modular suite of analysis tools. The physics engine rigorously models time-of-flight delays using high-order Lagrange interpolation, generates colored noise from arbitrary $1/f^\alpha$ power spectra, and supports user-defined modulation waveforms. This engine is coupled with a suite of interchangeable fitting algorithms. For high-throughput offline analysis, it features a highly-optimized, parallelized NLS implementation. For real-time state tracking, it provides a choice of time-domain Extended Kalman Filters (EKFs) with different process models to best match system dynamics. The package's object-oriented architecture facilitates the intuitive composition of virtual instruments from `laser' and `interferometer' objects. This is complemented by a high-throughput experimentation framework designed to automate large-scale parameter sweeps and Monte Carlo studies. Together, these features make \dfmk{} a powerful tool for optimizing instrument designs, investigating systematic errors, and developing novel signal processing techniques.

\subsection{Comparison to Existing Software}
\label{sec:comparison}

While \dfmk{}, to the best of our knowledge, is the first open-source package designed for end-to-end simulation and processing of DFMI systems, it is important to position it relative to existing tools in the broader field of computational interferometry. The current software landscape can be categorized into general-purpose interferometer simulators, commercial optical design packages, and specialized signal processing libraries, none of which provides an integrated solution for DFMI research.

\texttt{Finesse}~\cite{Finesse} is the state-of-the-art tool for frequency-domain simulation of complex, steady-state optical systems. It is unparalleled for modeling the optical cavities and advanced, coupled interferometer topologies found in gravitational wave detectors. The core challenge in DFMI, however, is the analysis of time-domain signals generated by a dynamically modulated laser. \texttt{Finesse}'s frequency-domain approach is not designed to model the crucial, time-varying time-of-flight effects and arbitrary modulation waveforms. Furthermore, it does not include the specialized parameter estimation algorithms, such as an NLS fitter or EKF, that are essential for a full end-to-end analysis. \dfmk{} is specifically designed to address this gap by providing a high-fidelity, time-domain physics engine directly coupled with a suite of appropriate signal processing back-ends.

Commercial optical design packages, such as Zemax OpticStudio~\cite{Zemax} and Code V~\cite{CodeV}, are industry standards for designing and tolerancing the physical layout of optical systems. While powerful for modeling geometric effects and aberrations, they are not specialized for the dynamic signal simulation and parameter estimation that define a DFMI measurement. A complete system performance analysis requires not only the optical model but also the injection of realistic, colored noise sources, the simulation of the resulting time-series, and the application of tailored demodulation algorithms. \dfmk{} provides this entire pipeline within a single, open-source framework, enabling an integrated approach where the impact of both optical design choices and signal processing strategies can be studied concurrently.

Finally, specialized libraries like \texttt{PyTDI}~\cite{pytdi} provide essential, high-performance algorithms for specific problems, such as the time-delay interferometry (TDI) processing required for the LISA mission~\cite{Tinto2005, Bayle2023}. Indeed, the robust Lagrange interpolation function used in \dfmk{} was adapted from the excellent, validated implementation in \texttt{PyTDI}. However, such libraries provide specific DSP components rather than a complete simulation framework. They do not include a physics engine to generate initial signals, models for various noise sources, or parameter estimation routines tailored to the DFMI signal model. \dfmk{} integrates this vital time-shifting capability into a comprehensive system that models the entire cause-and-effect chain of a DFMI instrument, from laser modulation to final parameter readout, thereby providing a unique and necessary tool for the field.

\subsection{Structure of the Paper}

This paper is structured as follows. In Section~\ref{sec:theory}, we review the theoretical signal model for a DFMI system. Section~\ref{sec:structure} provides a high-level overview of the \dfmk{} library's architecture. We then dedicate the subsequent sections to a detailed exploration of the framework's core components. Section~\ref{sec:readout} presents the parameter estimation engine, detailing the implementation strategies and performance of the NLS and EKF readout algorithms. The high-fidelity physics engine, with its focus on modeling dynamic interferometers and colored noise, is described in Section~\ref{sec:physics}. The framework for automating large-scale computational studies is presented in Section~\ref{sec:experiments}. We demonstrate the power of this framework in Section~\ref{sec:application} with an illustrative study of systematic errors due to modulation non-linearity, before summarizing our work and discussing future directions in Section~\ref{sec:summary-and-outlook}.

\section{Theory}
\label{sec:theory}

In this section, we develop the theoretical model of the signal generated by an unequal-arm interferometer under deep frequency modulation. We begin with a general formulation that accommodates arbitrary modulation waveforms and dynamic path length changes, which forms the basis of the \dfmk{} physics engine. We then specialize this model to the ideal case of sinusoidal frequency modulation, which is the foundation for the DFMI readout algorithms. Finally, we analyze the spectral content of this ideal signal, revealing how the interferometer's physical parameters are encoded in its harmonic structure.

\subsection{General Interferometric Signal Formulation}
\label{sec:general_signal}

\begin{figure}
\centering
\includegraphics[width=\columnwidth]{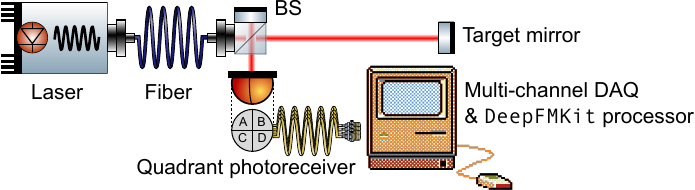}
\caption{Example DFMI optical configuration. A laser undergoing strong sinusoidal frequency modulation is injected into a Michelson interferometer. A beam splitter (BS) divides the incoming light and recombines the beams returning from the two interferometer arms. The reference arm is integrated within the BS assembly via a high-reflectivity mirror, while the measurement arm reflects from a target mirror. The recombined beam is directed to a quadrant photoreceiver. The photocurrents are converted to voltages by transimpedance amplifiers and digitized. The resulting signals are processed by \dfmk{} to provide high-resolution displacement, angular, and absolute range measurements of the target mirror.}
\label{fig:setup}
\end{figure}

We consider a single-frequency laser whose instantaneous angular frequency, $\omega(t)$, is deliberately modulated around a central carrier frequency, $\omega_0$. The total phase of the laser's electric field at time $t$, relative to its phase at $t=0$, is given by the integral of its instantaneous frequency:
\begin{equation}
\phi_{\text{laser}}(t) = \int_0^t \omega(t') dt' = \omega_0 t + \phi_{\text{mod}}(t),
\end{equation}
where $\phi_{\text{mod}}(t)$ is the phase modulation component. This laser beam is injected into a two-beam interferometer, such as the Michelson configuration shown in Figure~\ref{fig:setup}. The beam is split and travels along a reference arm and a measurement arm before the two are recombined.

Let $l_r(t)$ and $l_m(t)$ represent the time-dependent optical path lengths of the reference and measurement arms, respectively. The light traversing these paths experiences corresponding time-of-flight delays, $\tau_r(t) = l_r(t)/c$ and $\tau_m(t) = l_m(t)/c$, where $c$ is the speed of light. The electric fields from each arm, which arrive at the photodetector at time $t$, must have originated from the laser source at the earlier times $t-\tau_r(t)$ and $t-\tau_m(t)$. The superposition of these two fields creates an interference pattern. The optical power at the detector, and consequently the output voltage signal $v(t)$ after transimpedance amplification, is proportional to the cosine of the phase difference between the two recombining beams:
\begin{equation}
v(t) = A \left[ 1 + k \cos(\Delta\Phi_{\text{total}}(t)) \right],
\label{eq:voltage_general}
\end{equation}
where $A$ is a term proportional to the mean optical power at the photodetector, and $k$ is the fringe visibility, or contrast, of the interferometer. The total phase difference $\Delta\Phi_{\text{total}}(t)$ is the central quantity of interest and is given by the exact expression:
\begin{equation}
\Delta\Phi_{\text{total}}(t) = \phi_{\text{laser}}(t-\tau_r(t)) - \phi_{\text{laser}}(t-\tau_m(t)).
\label{eq:phase_diff_general}
\end{equation}
This general expression captures the complete physics of the interferometric process, including all time-of-flight and dynamic effects. It forms the basis of the high-fidelity physics engine implemented in \dfmk{}.

\subsection{Decomposition of the Phase Signal}

To better understand the physical origin of the signal, it is instructive to decompose the total phase difference. Substituting the expression for $\phi_{\text{laser}}(t)$ separates $\Delta\Phi_{\text{total}}(t)$ into two distinct terms:
\begin{align}
\Delta\Phi_{\text{total}}(t) = & \underbrace{\omega_0 (\tau_m(t) - \tau_r(t))}_{\text{carrier phase}} \nonumber \\ 
& + \underbrace{\left( \phi_{\text{mod}}(t-\tau_r(t)) - \phi_{\text{mod}}(t-\tau_m(t)) \right)}_{\text{modulation phase difference}}.
\label{eq:phase_decomposition}
\end{align}
The first term, the carrier phase, represents the conventional interferometric phase shift. Defining the optical path difference (OPD) as $\Delta l(t) = l_{r}(t) - l_{m}(t)$, this term can be written as $\omega_0 \Delta l(t)/c$. It contains the primary displacement information.

The second term, the modulation phase difference, is the unique signature of DFMI. It arises from the interference of the laser's own phase modulation with a delayed version of itself. The structure of this term depends on the specific waveform of $\phi_{\text{mod}}(t)$ and the time delays, and it is this component that creates the rich harmonic content of the DFMI signal.

\subsection{The Ideal DFMI Signal Model}

\begin{figure}
\centering
\includegraphics[width=\columnwidth]{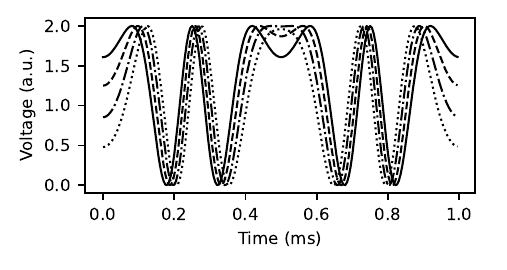}
\caption{Simulated voltage signals for the interferometer depicted in Figure~\ref{fig:setup}. The configuration has an OPD of $\Delta l = 5\,\rm cm$ and is injected with a laser modulated by $6.87\,\rm GHz$ at $1\,\rm kHz$, resulting in an effective modulation depth of $m = 7.2\,\rm rad$. The quadrant photoreceiver is sampled at 200\,kHz for one modulation cycle ($1/f_m = 1\,\rm ms$), producing the four time series shown.}
\label{fig:signals}
\end{figure}

DFMI relies on applying a high-purity sinusoidal frequency modulation. In the ideal case, the instantaneous frequency deviation from the carrier is:
\begin{equation}
f_{\text{mod}}(t) = \Delta f \cos(\omega_m t + \psi),
\label{eq:f_mod}
\end{equation}
where $\Delta f$ is the modulation or ``tuning'' amplitude, $\omega_m = 2 \pi f_m$ is the modulation angular frequency, and $\psi$ is the modulation phase offset. The corresponding phase modulation is the time integral of the angular frequency modulation, $2\pi f_{\text{mod}}(t)$:
\begin{equation}
\phi_{\text{mod}}(t) = \frac{2\pi\Delta f}{\omega_m} \sin(\omega_m t + \psi) = \frac{\Delta f}{f_m} \sin(\omega_m t + \psi).
\label{eq:phi_mod}
\end{equation}
In any practical implementation, the modulation frequency is low compared to the inverse light travel time ($\omega_m \tau \ll 1$). This allows the modulation phase difference term to be approximated using a first-order Taylor expansion~\cite{Dovale2025}:
\begin{equation}
\phi_{\text{mod}}(t) - \phi_{\text{mod}}(t-\tau) \approx \tau \frac{d}{dt}\phi_{\text{mod}}(t) = 2\pi\Delta f \tau \cos(\omega_m t + \psi).
\end{equation}
This approximation enables the definition of two key length-encoding parameters. The first is the effective modulation depth, $m$, which is unambiguously related to the OPD, $\Delta l = c\tau$:
\begin{equation}
m = 2\pi \Delta f \tau = \frac{2\pi \Delta f \Delta l}{c}.
\end{equation}
The second is the interferometric phase, $\Phi$, which contains the conventional $2\pi$ ambiguity:
\begin{equation}
\Phi = \omega_0 \tau = \frac{2\pi f_0 \Delta l}{c}.
\end{equation}
Combining these definitions, the total phase difference in this idealized model becomes $\Delta\Phi_{\text{total}}(t) = \Phi + m\cos(\omega_m t + \psi)$. The resulting voltage signal, depicted in Figure~\ref{fig:signals}, is given by the well-known expression:
\begin{equation}
v(t) = A \left[1 + k \cdot \cos\left(\Phi + m \cos \left( \omega_m t + \psi \right) \right)\right].
\label{eq:volt_signal}
\end{equation}
The time-varying part of this signal is periodic with frequency $f_m$ and depends parametrically on the state vector to be estimated. Grouping the AC amplitude terms into $C=Ak$, this vector is $\mathbf{x} = (C, m, \Phi, \psi)$. The parameters in this model correspond directly to the attributes of the configuration classes within \dfmk{}: 
\begin{itemize}
	\item The laser's modulation parameters ($f_m$, $\Delta f$, $\psi$) are defined by \texttt{LaserConfig.fm}, \texttt{LaserConfig.df}, and \texttt{LaserConfig.psi}.
	\item The signal's DC offset $A$ and AC amplitude $C$ are determined by the laser power, represented by \texttt{LaserConfig.amp}, and the interferometer's fringe visibility $k$, given by \texttt{IfoConfig.visibility}.
	\item The OPD $\Delta l$ is defined by the interferometer's arm lengths \texttt{IfoConfig.ref\_arml} and \texttt{IfoConfig.meas\_arml}, plus any prescribed motion or noise, and an arbitrary phase offset \texttt{IfoConfig.phi}.
\end{itemize}

\subsection{The Signal in the Frequency Domain}

To understand how the parameters are encoded in the signal's harmonic structure, we analyze the spectral content of the ideal AC component, $v_{\text{AC}}(t) = C \cos(\Phi + m \cos(\omega_m t + \psi))$. Using the angle-sum identity for cosine, this becomes:
\begin{equation}
v_{\text{AC}}(t) = C \left[ \cos(\Phi)\cos(m\cos\theta) - \sin(\Phi)\sin(m\cos\theta) \right],
\label{eq:v_ac_expanded}
\end{equation}
where for brevity we define $\theta = \omega_m t + \psi$. The terms containing $m$ can be expanded into an infinite series of harmonics using the Jacobi-Anger identities~\cite{NIST:DLMF}:
\begin{align}
\cos(m\cos\theta) &= J_0(m) + 2 \sum_{n=2, \text{even}}^\infty (-1)^{n/2} J_n(m) \cos(n\theta), \label{eq:ja_cos}\\
\sin(m\cos\theta) &= 2 \sum_{n=1, \text{odd}}^\infty (-1)^{(n-1)/2} J_n(m) \cos(n\theta), \label{eq:ja_sin}
\end{align}
where $J_n(m)$ is the Bessel function of the first kind of order $n$. Substituting these expansions into Eq.~\eqref{eq:v_ac_expanded} reveals that the signal is a sum of harmonics of the modulation frequency $f_m$.

The complex amplitude of each harmonic, $\alpha_n$, can be extracted by projecting the signal onto reference cosine and sine waves to find the in-phase ($I_n$) and quadrature ($Q_n$) components:
\begin{align}
I_n(\mathbf{x}) &= \frac{2}{T} \int_{0}^{T} v_{\text{AC}}(t) \cos(n\omega_m t)\, dt \label{eq:I_integral} \\
Q_n(\mathbf{x}) &= \frac{2}{T} \int_{0}^{T} v_{\text{AC}}(t) \sin(n\omega_m t)\, dt, \label{eq:Q_integral}
\end{align}
where $T=1/f_m$ is the modulation period. Solving these integrals yields the analytical expressions for the harmonic amplitudes:
\begin{align}
I_n(\mathbf{x}) &= 2C \cdot J_n(m) \cos\left(\Phi + n\frac{\pi}{2}\right) \cos(n\psi) \label{eq:Iana} \\
Q_n(\mathbf{x}) &= -2C \cdot J_n(m) \cos\left(\Phi + n\frac{\pi}{2}\right) \sin(n\psi). \label{eq:Qana}
\end{align}
From these, we construct the complex harmonic amplitude $\alpha_n = I_n + iQ_n$, which has the compact form:
\begin{equation}
\alpha_n(\mathbf{x}) = 2C \cdot J_n(m) \cos\left(\Phi + n\frac{\pi}{2}\right) e^{-in\psi}.
\label{eq:alpha_n}
\end{equation}
This equation clearly illustrates the unique role of each parameter. The modulation depth $m$ governs the distribution of power among the harmonics via the Bessel functions $J_n(m)$. The modulation phase $\psi$ introduces a differential phase rotation, where the $n$-th harmonic is rotated by an angle of $-n\psi$. Finally, the interferometric phase $\Phi$ acts as a real-valued scaling factor that modulates the even and odd harmonics differently through the term $\cos(\Phi+n\pi/2)$. These distinct signatures form the basis of the frequency-domain NLS readout algorithm.

\section{Software Architecture}
\label{sec:structure}

\dfmk{} is built on a modular, object-oriented architecture in Python~\cite{python-language-reference}. The design philosophy is to separate the core concerns of data representation, physics simulation, signal processing, and parameter estimation into distinct, interchangeable components. This structure not only promotes code clarity and maintainability but also provides researchers with the flexibility to extend the framework or substitute its components with their own implementations. At a high level, a main controller class, \texttt{DeepFrame}, orchestrates the interactions between these modules, as demonstrated in the typical workflow shown in Figure~\ref{fig:workflow}. This section details the architecture and function of each key component of the library.

\subsection{Core Controller: \texttt{DeepFrame}}
\label{sec:core}

The central user-facing component of the library is the \texttt{DeepFrame} class, found in \texttt{core.py}. This class acts as a high-level controller that manages the state of an entire analysis session. It maintains dictionaries of the various objects that constitute an experiment:
\begin{itemize}
    \item \texttt{sims}: A dictionary of \texttt{SimConfig} instances, each defining the complete physical configuration for a simulation channel.
    \item \texttt{raws}: A dictionary of \texttt{RawData} instances, which hold raw time-series data, either loaded from a file or generated by the physics engine.
    \item \texttt{fits}: A dictionary of \texttt{FitData} instances, containing the time-series results from a parameter estimation algorithm.
\end{itemize}
The primary methods of \texttt{DeepFrame}, such as \texttt{.simulate()} and \texttt{.fit()}, provide a simple interface to the more complex underlying modules. For example, calling \texttt{df.simulate()} instantiates the \texttt{SignalGenerator} physics engine, runs a simulation based on a configuration from \texttt{df.sims}, and stores the resulting time series in a new \texttt{RawData} object within \texttt{df.raws}. Similarly, calling \texttt{df.fit()} uses one of the available readout algorithms from \texttt{fitters.py} to process a \texttt{RawData} object, generating a corresponding \texttt{FitData} object. This orchestration simplifies the user workflow, allowing researchers to focus on experimental design rather than implementation details.

\begin{figure}
\centering
\begin{python}
import deepfmkit.core as dfm
import matplotlib.pyplot as plt
# --- 1. Define the interferometer ---
ifo = dfm.IfoConfig()
ifo.ref_arml = 0.10 # Reference armlength (m)
ifo.meas_arml = 0.15 # Measurement armlength (m)
# --- 2. Define the laser source ---
las = dfm.LaserConfig()
las.fm = 1e3 # FM frequency (Hz)
las.set_df_for_m(ifo, 6.) # Target m (rad)
# --- 3. Compose the main channel ---
sim = dfm.SimConfig("ch0",
    laser_config=las,
    ifo_config=ifo,
    f_samp=200e3) # Acquisition frequency (Hz) 
# --- 4. Instantiate DeepFrame ---
df = dfm.DeepFrame(sim_config=sim)
# --- 5. Simulate ---
df.simulate("ch0",
    n_seconds=10) # Simulation length (s)
# --- 6. Run readout algorithm ---
df.fit("ch0")
# --- 7. Plot readout results ---
axes = df.plot(which=["phi", "m", "ssq"])
plt.show()
\end{python}
\caption{\dfmk{}'s basic ``simulate-and-analyze'' workflow.}
\label{fig:workflow}
\end{figure}

\subsection{Data Structures: \texttt{data.py}}
\label{sec:data}

To ensure data consistency and portability throughout the framework, we define two primary data structures using Python's \texttt{dataclasses}. These objects act as standardized containers for both data and its associated metadata.
\begin{itemize}
    \item {\texttt{RawData}}: This class encapsulates a raw, single-channel time series, whether loaded from a file or generated by the physics engine. The data is stored in a \texttt{pandas.DataFrame}~\cite{McKinney2010}. Crucially, this object also holds essential metadata, such as the sampling frequency (\texttt{f\_samp}) and modulation frequency (\texttt{fm}). If the data was generated by the internal physics engine, the object maintains a link to the simulation configuration and stores the ground-truth time series for all noise sources and the ideal phase signal, enabling direct comparison for benchmarking analysis algorithms.
    \item {\texttt{FitData}}: This class stores the results of a parameter estimation routine. It contains the time series of the fitted parameters (e.g., $C$, $m$, $\Phi, \psi$), the sum-of-squares error (\texttt{ssq}), and metadata about the fit itself, such as the number of modulation cycles per fit buffer (\texttt{n}) and the resulting fit data rate (\texttt{fs}). This structure facilitates the plotting, comparison, and storage of analysis results. The class also includes a method for loading fit data from files, such as those recorded by a DFMI processor in a laboratory experiment.
\end{itemize}

\subsection{Physics Engine: \texttt{physics.py}}
\label{sec:physics}

The generation of realistic DFMI signals is handled by the physics engine, implemented in the \texttt{physics.py} module. The design follows the Composition software pattern, allowing for the intuitive construction of complex experimental setups from simpler, reusable components.

\paragraph{Configuration Objects}
The physical system is defined by three configuration classes that are composed together: \texttt{LaserConfig}, \texttt{IfoConfig}, and \texttt{SimConfig}.
\begin{itemize}
    \item \texttt{LaserConfig} encapsulates all properties of the light source, including its \texttt{wavelength}, a term representing optical power (\texttt{amp}), the modulation frequency (\texttt{fm}), and tuning amplitude (\texttt{df}). A key feature is the \texttt{waveform\_func} attribute, which accepts any Python callable. This allows users to define arbitrary laser modulation waveforms beyond the ideal sinusoid, with a library of common functions provided in \texttt{waveforms.py}. Users can also inject various noise terms, including laser frequency noise (\texttt{f\_n}), tuning amplitude noise (\texttt{df\_n}), and relative intensity noise (\texttt{r\_n}).
    \item \texttt{IfoConfig} defines the properties of the interferometer, such as its static arm lengths (\texttt{ref\_arml} and \texttt{meas\_arml}) and fringe \texttt{visibility}. It also allows for the inclusion of prescribed sinusoidal motion (via \texttt{arml\_mod\_f} and \texttt{arml\_mod\_amp}) as well as stochastic noise sources like OPD noise (\texttt{arml\_n}) and detector noise (\texttt{s\_n}).
    \item \texttt{SimConfig} represents a complete measurement channel by composing one \texttt{LaserConfig} with one \texttt{IfoConfig}. This compositional structure is powerful; for example, a single \texttt{LaserConfig} object can be shared between two different \texttt{IfoConfig} objects to naturally model a setup where one laser feeds both a primary and a reference interferometer. 
\end{itemize}
These classes include validation logic in their \texttt{\_\_post\_init\_\_} methods to prevent the creation of physically inconsistent configurations.

\paragraph{Signal Generation}
The \texttt{SignalGenerator} class is the core physics engine. Its primary method, \texttt{.generate()}, takes one or more \texttt{SimConfig} objects and produces the corresponding \texttt{RawData} objects. The internal \texttt{\_run\_physics\_simulation} method implements the full physical model from Section~\ref{sec:theory}, using the \texttt{dsp.timeshift} function to apply precise, time-varying delays to the modulation waveform. This accurately captures effects arising from a dynamic time-of-flight delay. To mitigate boundary artifacts from the finite impulse response (FIR) filtering used in \texttt{timeshift}, the engine employs a robust ``pad-and-crop'' strategy: it simulates extra modulation cycles at the beginning and end of the time series and then crops the output to the requested length, ensuring the numerical validity of the entire final dataset.

\begin{figure}[t!]
\centering
\includegraphics[width=\columnwidth]{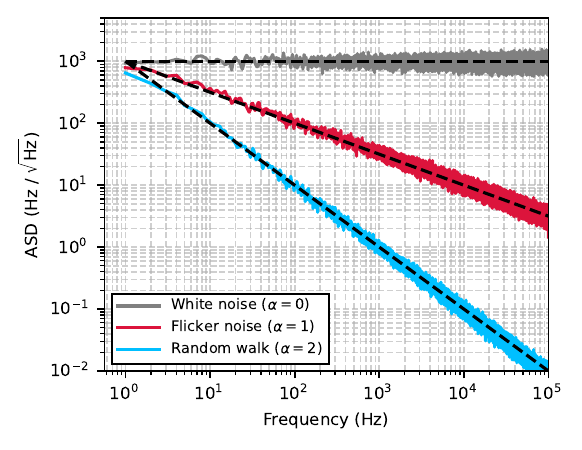}
\caption{Validation of the colored noise generation engine. The plot shows the amplitude spectral densities (ASDs) of generated time series for three different noise colors: white noise ($\alpha=0$), flicker noise ($\alpha=1$), and random walk noise ($\alpha=2$). The solid colored lines show the ASDs of the generated noise, which closely match the theoretical $1/f^{\alpha/2}$ slopes (dashed black lines), confirming the fidelity of the noise model.}
\label{fig:noise_validation}
\end{figure}

\subsection{Specialized Support Modules: \texttt{dsp.py} and \texttt{noise.py}}
\label{sec:dsp_noise}

\paragraph{Digital Signal Processing (\texttt{dsp.py})}
This module provides specialized signal processing functions. Its most critical component is the \texttt{timeshift} function, adopted from the \texttt{PyTDI} project~\cite{pytdi}, which implements a high-order Lagrange interpolator for applying fractional time delays to a signal. This function is essential for the high-fidelity physics model, where the time-of-flight in each interferometer arm is a dynamic quantity. The implementation is fully vectorized using NumPy's \texttt{einsum} and stride tricks~\cite{Harris2020}, avoiding slow Python loops to achieve high performance even for time-varying shifts.

\paragraph{Noise Generation (\texttt{noise.py})}
All noise terms declared in the physical configuration objects \texttt{LaserConfig} and \texttt{IfoConfig} are accompanied by an associated ``\texttt{\_alpha}'' term (e.g., \texttt{f\_n\_alpha}) that describes the frequency scaling of its power spectral density (PSD):
\begin{equation}
	\text{PSD} \propto \frac{1}{f^{\alpha}}
\end{equation}
where $\alpha \in [0,2]$. For example, the user defines $\alpha = 0$ for white noise, $\alpha = 1$ for pink (flicker) noise, or $\alpha = 2$ for red (random walk) noise. The noise terms themselves (e.g., \texttt{f\_n}) conveniently describe the value of the amplitude spectral density (ASD) at 1\,Hz. Figure~\ref{fig:noise_validation} shows the resulting ASDs of noise streams generated by the noise engine with different $\alpha$ settings.

The framework's ability to simulate realistic noise is based on the \texttt{noise.py} module, which can generate time-series data with an arbitrary $1/f^\alpha$ power spectrum. The implementation follows the numerically stable method of Plaszczynski~\cite{PLASZCZYNSKI2007}, which uses a cascade of first-order infinite impulse response (IIR) filters to shape white noise. To achieve high performance suitable for generating long time-series, the performance-critical filtering loop in the \texttt{alpha\_noise} class is Just-In-Time (JIT) compiled using Numba~\cite{Numba}, yielding execution speeds comparable to compiled languages while preserving the stability of the original algorithm.

\begin{figure*}[htpb!]
\centering
\includegraphics{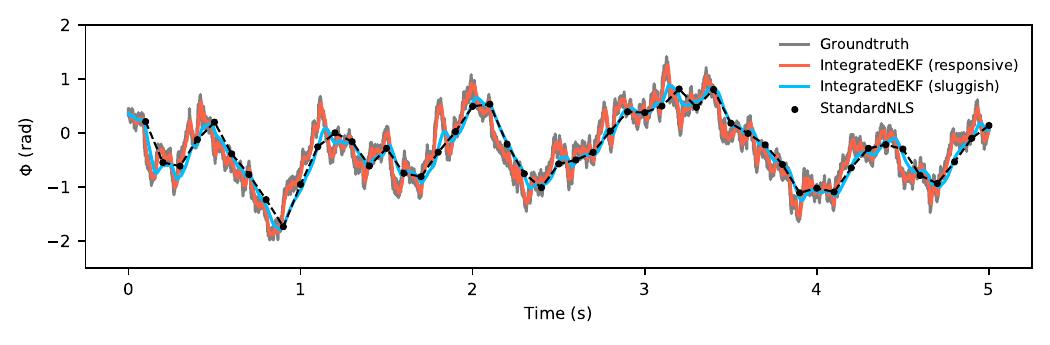}
\caption{Comparison of the \texttt{StandardNLS} and \texttt{IntegratedEKF} fitters in tracking a highly dynamic phase signal. A strong random walk phase noise (``Groundtruth'') was simulated by injecting a large amount of random walk laser frequency noise (100\,MHz/$\sqrt{\rm Hz}$ at 1\,Hz) into an interferometer model with a 20\,cm OPD. The EKF output is shown for two different tunings of the process noise covariance: a responsive filter (red) that closely tracks the high-frequency dynamics at the cost of higher estimate variance, and a more heavily filtered configuration (blue) that provides a smoother, but slightly sluggish, estimate. The batch-processing NLS fitter (black dots), configured to provide an estimate of the average phase over 100\,ms data buffers, shows similar lag as the sluggish EKF filter with low process noise covariance.}
\label{fig:readout_comparison}
\end{figure*}

\subsection{Parameter Estimation: \texttt{fitters.py}}
\label{sec:fitters}

Parameter estimation is managed by a suite of interchangeable ``fitter'' classes defined in \texttt{fitters.py}. This module is designed around the Strategy software pattern: a \texttt{BaseFitter} abstract class defines a common interface, and concrete classes like \texttt{StandardNLS}, \texttt{StandardEKF}, and \texttt{IntegratedEKF} provide specific implementations of different algorithms. This design allows a user to select a readout algorithm at runtime via the \texttt{df.fit(method='...')} call.

\subsubsection{NLS Fitter}
The \texttt{StandardNLS} class provides a high-performance, frequency-domain Non-Linear Least Squares fitter optimized for offline analysis. It operates on discrete data buffers and estimates the four primary DFMI parameters: the AC amplitude $C$, modulation depth $m$, interferometric phase $\Phi$, and modulation phase $\psi$. The underlying Levenberg-Marquardt algorithm and its performance optimizations, including vectorization and parallel execution, are implemented in the low-level \texttt{fit.py} module.

\subsubsection{EKF Fitters}
As a time-domain alternative for real-time state estimation, \dfmk{} provides two distinct Extended Kalman Filter implementations. Both operate directly on the raw signal samples and are designed for high-speed execution. The choice of filter allows users to select the process model that best matches their system's expected dynamics~\cite{Sarkka2013, Im2024}.

The first, \texttt{StandardEKF}, implements an EKF with a random walk process model. This is a common and robust choice for systems where parameter dynamics are unknown or expected to be stochastic. It tracks a 5-dimensional state vector containing the four primary DFMI parameters ($C, m, \Phi, \psi$) and the signal's DC offset $A$.

The second, \texttt{IntegratedEKF}, uses a more complex integrated random walk (or constant velocity) process model. This filter expands the state vector to 10 dimensions to include the rates of change for all five signal parameters. It is designed for systems where parameters are expected to exhibit a persistent, linear drift, as it can project this trend forward in its prediction step.

To briefly illustrate the complementary strengths of the batch-processing NLS fitter and the real-time EKF, we simulated their performance in tracking a highly dynamic signal. We inject a large amount of laser frequency noise in an unequal-arm interferometer with 20\,cm OPD, and use three differently-configured fitters to track the resulting highly dynamic interferometric phase (Figure~\ref{fig:readout_comparison}).

\subsection{High-Throughput Experiments: \texttt{experiments.py}}
\label{sec:experiments}
For systematic studies involving large parameter sweeps or Monte Carlo simulations, \dfmk{} provides a high-level experimentation framework in the \texttt{experiments.py} module. The \texttt{Experiment} class allows a user to declaratively define a multi-dimensional parameter space, including static parameters, swept axes, and stochastic variables.

A central feature of this design is the \texttt{ExperimentFactory} software pattern, implemented in \texttt{factories.py}. To ensure that user-defined logic for configuring physics objects can be safely serialized (``pickled'') and sent to worker processes for parallel execution, the user encapsulates this logic within a class that inherits from \texttt{ExperimentFactory}. An instance of this factory is then passed to the \texttt{Experiment} object. When the \texttt{.run()} method is called, it generates a complete list of all job configurations and distributes them to a pool of worker processes. Each worker executes the entire simulate-and-analyze pipeline for a single set of parameters. This architecture prevents nested parallelism, which can lead to performance degradation. and enables robust, large-scale computational experiments for thoroughly characterizing DFMI system performance.

\section{Parameter Estimation Engine}
\label{sec:readout}

\subsection{The Non-Linear Least Squares Fitter}
\label{sec:NLS}
The Non-linear Least Squares (NLS) fitter is a cornerstone of \dfmk{} for high-precision, very high-throughput, online or offline DFMI signal processing. The implementation, found in \texttt{fitters.py} and \texttt{fit.py}, is engineered for a combination of speed, robustness, and flexibility. It is a batch-processing algorithm that operates on discrete buffers of the raw time series, where each buffer contains an integer number of modulation cycles to ensure harmonic orthogonality.

\subsubsection{Core Algorithm}
The core of the NLS fitter is an in-house implementation of the Levenberg-Marquardt Algorithm (LMA)~\cite{Levenberg1944, Marquardt1963, More1978}, tailored specifically to the DFMI signal model described in Section~\ref{sec:theory}. The low-level, performance-critical implementation of the LMA resides in \texttt{fit.py} and was originally derived from an unpublished C implementation by Gerhard Heinzel~\cite{Heinzel2010, HeinzelLMA}. The algorithm's structure and key optimizations are detailed below.

\paragraph{The LMA Loop}
The LMA iteratively refines an initial guess for the parameter vector,
\begin{equation}
\mathbf{x} = (C, m, \Phi, \psi).	
\end{equation}
At each step, it calculates the sum-of-squared-residuals (SSQ) between the measured I/Q harmonic amplitudes and the model from Eq.~\eqref{eq:alpha_n}, as well as the Jacobian matrix $\mathbf{J}$ of the model with respect to the parameters. A key optimization in \dfmk{} is that the LMA loop calls the comprehensive \texttt{ssq\_jac\_grad} function (which computes SSQ, $\mathbf{J}^\mathrm{T}\mathbf{J}$, and the gradient $\mathbf{J}^\mathrm{T}\mathbf{r}$) only once per successful step. When testing different damping parameters $\lambda$ to find an optimal step, it uses a lightweight, vectorized \texttt{ssqf} function that calculates only the SSQ for a trial parameter set, thus avoiding redundant Jacobian calculations. The overall logic is captured in Algorithm~\ref{alg:lma}.

\begin{algorithm}[h!]
\caption{Optimized LMA loop in \texttt{fit.py}}
\label{alg:lma}
\begin{algorithmic}[1]
\STATE \textbf{Input:} Measured I/Q data $\bm{\alpha}$, initial guess $\mathbf{x}_0$
\STATE $\mathbf{x} \leftarrow \mathbf{x}_0$
\STATE SSQ, $\mathbf{A}$, $\mathbf{g} \leftarrow \text{ssq\_jac\_grad}(\bm{\alpha}, \mathbf{x})$ \COMMENT{$\mathbf{A}=\mathbf{J}^\mathrm{H}\mathbf{J}$, $\mathbf{g}=\mathbf{J}^\mathrm{H}\mathbf{r}$}
\FOR{$k=1$ \TO MAX\_ITER}
    \STATE $\text{improved} \leftarrow \text{false}$
    \FOR{$\lambda$ in $[0, 10^{-7}, \dots, 100]$}
        \STATE $\bm{\delta} \leftarrow \text{msolve}(\lambda, \mathbf{A}, \mathbf{g})$ \COMMENT{Solve damped equation}
        \STATE $\mathbf{x}_{\text{try}} \leftarrow \mathbf{x} + \bm{\delta}$
        \STATE $\text{SSQ}_{\text{try}} \leftarrow \text{ssqf}(\bm{\alpha}, \mathbf{x}_{\text{try}})$
        \IF{$\text{SSQ}_{\text{try}} < \text{SSQ}$}
            \STATE $\mathbf{x} \leftarrow \mathbf{x}_{\text{try}}$
            \STATE SSQ, $\mathbf{A}$, $\mathbf{g} \leftarrow \text{ssq\_jac\_grad}(\bm{\alpha}, \mathbf{x})$
            \STATE $\text{improved} \leftarrow \text{true}$
            \STATE \textbf{break} \COMMENT{Accept step and decrease $\lambda$}
        \ENDIF
    \ENDFOR
    \IF{\textbf{not} improved}
        \STATE \textbf{break} \COMMENT{No improvement found, converged}
    \ENDIF
    \IF{converged$(\mathbf{x}, \mathbf{x}_{\text{prev}}, \text{SSQ}, \text{SSQ}_{\text{prev}})$}
        \STATE \textbf{break}
    \ENDIF
\ENDFOR
\STATE \textbf{return} $\mathbf{x}$, SSQ
\end{algorithmic}
\end{algorithm}

\subsubsection{Robustness and Automated Initialization}
A well-known challenge for NLS algorithms is their sensitivity to the initial parameter guess, which can lead to convergence in a local minimum of the SSQ landscape. \dfmk{}'s NLS fitter incorporates a two-stage automated initialization strategy to enhance robustness against poor starting values.

\begin{enumerate}
    \item \textbf{Smart $\psi$ initialization}: Before the main LMA commences, the fitter can perform an optional one-dimensional scan over the modulation phase, $\psi$. Because an incorrect initial $\psi$ is a common cause of poor convergence, this pre-fit step robustly finds a good starting point by minimizing the SSQ of a trial fit on the first data buffer, significantly improving the likelihood of global convergence.

    \item \textbf{Grid-search on failure}: If a primary fit results in an SSQ value above a quality threshold, the fitter presumes it has converged to a local minimum. In this event, it triggers a fallback routine that performs a coarse grid search over the modulation depth, $m$. For each trial $m$, it analytically estimates the corresponding optimal $C$ and $\Phi$. A full LMA fit is then initiated from the most promising point on this grid. This automated retry mechanism makes the fitter resilient to poor initial conditions, particularly in low signal-to-noise ratio scenarios.
\end{enumerate}

\subsubsection{Performance Optimizations}
The high throughput of the NLS fitter is achieved through a combination of vectorization, parallel execution, and algorithmic optimizations.

\paragraph{Vectorization}
The primary source of the fitter's high performance is the extensive use of vectorization via the NumPy library~\cite{Harris2020}. The core \texttt{ssq\_jac\_grad} and \texttt{ssqf} functions compute the model's predicted I/Q values and their derivatives for all $N$ harmonics simultaneously, avoiding slow Python loops. Operations such as evaluating Bessel functions (\texttt{scipy.special.jv}~\cite{SciPy}), trigonometric functions, and the final matrix products ($\mathbf{J}^\mathrm{T}\mathbf{J}$ and $\mathbf{J}^\mathrm{T}\mathbf{r}$) are performed on entire NumPy arrays. This approach leverages the highly optimized, compiled code underlying NumPy, resulting in a dramatic speedup compared to a naive, iterative implementation.

\paragraph{Parallel Execution and Warm-Starts}
Fitting a long time series involves processing hundreds or thousands of independent data buffers, a task that is embarrassingly parallel. The \texttt{StandardNLS} class orchestrates this by splitting the full dataset into chunks and distributing them to a pool of worker processes using Python's \texttt{multiprocessing} library. Each worker process executes the LMA on its assigned data chunks. Furthermore, a "warm-start" optimization is employed: the converged parameter vector from fitting buffer $i$ is used as the initial guess for buffer $i+1$. Because physical parameters typically change slowly between adjacent buffers, this strategy dramatically reduces the number of LMA iterations required for convergence on subsequent buffers, further enhancing overall throughput.

\subsection{The Extended Kalman Filter}
\label{sec:EKF}

As a complementary, time-domain alternative to the batch-processing NLS fitter, \dfmk{} provides two distinct Extended Kalman Filter (EKF) implementations. EKFs are designed for real-time state estimation, providing an updated estimate of the interferometer's parameters with each incoming data sample. This makes them well-suited for tracking dynamic systems where parameters may be changing rapidly.

The core of any EKF is the recursive predict-update cycle, summarized in Algorithm~\ref{alg:ekf}. The key difference between the two EKF implementations in \dfmk{} lies in the underlying state-space model chosen to represent the system's dynamics.

\begin{algorithm}[h!]
\caption{The general Extended Kalman Filter (EKF) cycle. The matrices $\mathbf{F}$ and $\mathbf{H}_k$, and the vector $\mathbf{x}_k$ are defined by the specific state-space model.}
\label{alg:ekf}
\begin{algorithmic}[1]
\STATE \textbf{Initialize:} state estimate $\hat{\mathbf{x}}_{0|0}$, covariance $\mathbf{P}_{0|0}$
\FOR{$k=1$ \TO $N_{\text{samples}}$}
\STATE \textbf{Phase 1.\ Predict Step}
\STATE $\hat{\mathbf{x}}_{k|k-1} \leftarrow \mathbf{F} \hat{\mathbf{x}}_{k-1|k-1}$ \COMMENT{Project state ahead}
\STATE $\mathbf{P}_{k|k-1} \leftarrow \mathbf{F} \mathbf{P}_{k-1|k-1} \mathbf{F}^\top + \mathbf{Q}$ \COMMENT{Project error covariance}
\STATE \textbf{Phase 2.\ Update Step}
\STATE $\mathbf{H}_k \leftarrow \frac{\partial h}{\partial \mathbf{x}} \Big|_{\hat{\mathbf{x}}_{k|k-1}}$ \COMMENT{Linearize measurement model}
\STATE $\tilde{y}_k \leftarrow z_k - h(\hat{\mathbf{x}}_{k|k-1})$ \COMMENT{Compute innovation}
\STATE $\mathbf{S}_k \leftarrow \mathbf{H}_k \mathbf{P}_{k|k-1} \mathbf{H}_k^\top + R$ \COMMENT{Compute innov.\ covariance}
\STATE $\mathbf{K}_k \leftarrow \mathbf{P}_{k|k-1} \mathbf{H}_k^\top \mathbf{S}_k^{-1}$ \COMMENT{Compute optimal Kalman gain}
\STATE $\hat{\mathbf{x}}_{k|k} \leftarrow \hat{\mathbf{x}}_{k|k-1} + \mathbf{K}_k \tilde{y}_k$ \COMMENT{Update state estimate}
\STATE $\mathbf{P}_{k|k} \leftarrow (\mathbf{I} - \mathbf{K}_k \mathbf{H}_k)\mathbf{P}_{k|k-1}$ \COMMENT{Update error covariance}
\ENDFOR
\end{algorithmic}
\end{algorithm}

\subsubsection{Model 1: Random Walk EKF (\texttt{StandardEKF})}
The first implementation, \texttt{StandardEKF}, employs a simple and robust random walk (RW) process model. This is a common choice for systems where the exact parameter dynamics are unknown or are expected to be stochastic.

\paragraph{State-Space Model}
The state vector $\mathbf{x}_k$ is 5-dimensional, containing the four core DFMI parameters ($C, m, \Phi, \psi$) and the DC offset ($A$):
\begin{equation}
    \mathbf{x}_k = \begin{bmatrix}  C_k, & m_k, & \Phi_k, & \psi_k, & A_k \end{bmatrix}^\top.
\end{equation}
The RW model assumes the state at time $k$ is equal to the state at time $k-1$ plus zero-mean Gaussian process noise; the state transition matrix $\mathbf{F}$ is therefore the identity matrix, $\mathbf{I}$. The measurement function $h(\mathbf{x}_k)$ relates the state to the raw voltage measurement $z_k$:
\begin{equation}
    h(\mathbf{x}_k) = C_k \cos\left(\Phi_k + m_k \cos(\omega_m t_k + \psi_k)\right) + A_k.
\end{equation}
The filter linearizes this function at each step by computing its Jacobian, a $1 \times 5$ row vector $\mathbf{H}_k$. This model's primary advantage is its simplicity and stability, making it effective for tracking quasi-static parameters or those subject to random jitter.

\subsubsection{Model 2: Integrated Random Walk EKF\\(\texttt{IntegratedEKF})}
The second implementation, \texttt{IntegratedEKF}, uses a more complex integrated random walk (IRW), or constant velocity, process model. This model is designed for systems where parameters are expected to exhibit a persistent, linear drift over time.

\paragraph{State-Space Model}
To model parameter velocity, the state vector is expanded to 10 dimensions to include the rates of change for all five signal parameters:
\begin{equation}
    \mathbf{x}_k = \left[ C_k, \dot{C}_k, m_k, \dot{m}_k, \Phi_k, \dot{\Phi}_k, \psi_k, \dot{\psi}_k, A_k, \dot{A}_k \right]^\top.
\end{equation}
The state transition matrix $\mathbf{F}$ is no longer the identity matrix but is instead a $10 \times 10$ block-diagonal matrix that projects the state forward based on the current rates. For a time step $\Delta t$, it is composed of five identical kinematic blocks:
\begin{equation}
\mathbf{F} = \text{diag}\left( \mathbf{F}_{\text{block}}, \dots, \mathbf{F}_{\text{block}} \right), \quad \text{where} \quad \mathbf{F}_{\text{block}} = \begin{bmatrix} 1 & \Delta t \\ 0 & 1 \end{bmatrix}.
\end{equation}
The measurement function $h(\mathbf{x}_k)$ is identical to that of the RW EKF, but its Jacobian $\mathbf{H}_k$ is now a sparse $1 \times 10$ row vector, as the measurement depends only on the parameter values, not their rates. This model can more accurately track parameters with constant drift but requires more careful tuning of its larger process noise covariance matrix, $\mathbf{Q}$.

\subsubsection{Operational Characteristics}
The performance of both EKF models is highly dependent on the tuning of their respective covariance matrices: $\mathbf{Q}$ (process noise) and $R$ (measurement noise). The measurement noise variance, $R$, can often be estimated directly from the raw signal's variance. The process noise, $\mathbf{Q}$, is a critical user-tunable parameter that must be set to match the expected system dynamics. A larger $\mathbf{Q}$ allows the filter to track faster parameter changes at the cost of higher estimate variance, whereas a smaller $\mathbf{Q}$ results in smoother estimates that may lag behind rapid dynamics. The choice between the two EKF models allows a researcher to select the process model that best reflects the physical reality of their instrument.

\subsection{High-Performance Implementation in Python}
\label{sec:performance}

A primary goal of \dfmk{} is to provide not only physically accurate models but also high-performance tools capable of processing large experimental datasets efficiently. Achieving this in a high-level language like Python requires deliberate optimization strategies. This section details the specific techniques used to maximize the computational performance of the NLS and EKF fitters, leading to the throughputs reported in Table~\ref{tab:fitter_performance}.

\subsubsection{NLS Fitter: Vectorization and Parallelism}
The high throughput of the \texttt{StandardNLS} fitter is achieved through a combination of vectorization, parallel processing, and an algorithmic warm-start optimization.

\paragraph{Vectorization}
The core mathematical operations of the NLS fitter, particularly the calculation of the Jacobian and the sum-of-squared-residuals in the \texttt{ssq\_jac\_grad} function, are fully vectorized using NumPy~\cite{Harris2020}. Instead of iterating over each harmonic in a Python loop, the model equations are evaluated for all harmonics simultaneously using array operations. This approach delegates computationally intensive loops to NumPy's underlying compiled libraries, minimizing the overhead of the Python interpreter and dramatically accelerating the computation for each fit.

\paragraph{Parallel Execution}
The task of fitting thousands of independent data buffers is embarrassingly parallel. The \texttt{StandardNLS} class leverages this by using Python's \texttt{multiprocessing} library to distribute chunks of data buffers across all available CPU cores. Each core executes the fitting algorithm on its assigned chunk independently. This data-parallel approach allows the total processing time to scale nearly linearly with the number of available cores, making it possible to analyze very large datasets in a fraction of the time required for sequential processing.

\paragraph{Warm-Start Optimization}
In addition to these computational strategies, the fitter employs a "warm-start" algorithmic optimization for processing sequential data buffers. The converged parameter vector from fitting buffer $i$ is used as the initial guess for buffer $i+1$. For signals with slowly varying parameters, this initial guess is often extremely close to the true solution. As a result, the Levenberg-Marquardt algorithm typically converges in very few iterations (often just one or two) for all but the first buffer in a contiguous data block. This allows the fitter to bypass the expensive iterative search for the majority of the dataset, contributing significantly to the high throughputs observed in benchmarks.

\subsubsection{EKF Fitters: Just-In-Time (JIT) Compilation}
The recursive nature of the EKF algorithm means its performance is limited by the speed of its sample-by-sample update loop, which would be a significant bottleneck in a standard Python implementation.

To overcome this, \dfmk{} isolates the core predict-update cycle of both the \texttt{StandardEKF} and \texttt{IntegratedEKF} into dedicated helper functions. These functions are decorated with \texttt{@jit(nopython=True)} from the Numba library~\cite{Numba}. The JIT compiler translates the Python code for these critical loops into highly optimized machine code the first time they are called. All subsequent calls execute this fast, compiled version, bypassing the Python interpreter entirely. This technique provides performance comparable to a compiled language for the most computationally intensive part of the algorithm.

The performance difference between the two EKF implementations seen in Table~\ref{tab:fitter_performance} is a direct consequence of their differing model complexity. The 10-dimensional IRW EKF (\texttt{IntegratedEKF}) performs more floating-point operations per sample than the 5-dimensional version due to its larger state vector (e.g., $10 \times 10$ vs. $5 \times 5$ matrix multiplications) and the non-identity state transition matrix multiplication required in the predict step.

\subsection{Performance Benchmarks}
To quantify the performance of these optimizations, we benchmarked all fitters on modern laptop workstations. The task was to process a 100-second raw data stream sampled at 200\,kHz, representing 20 million data points. The results, summarized in Table~\ref{tab:fitter_performance}, highlight the effectiveness of the different optimization strategies.

\begin{table}[h!]
\centering
\caption{Fitter performance benchmarks. Throughput is measured in thousands of raw input samples processed per second (kS/s).}
\label{tab:fitter_performance}
\small
\begin{tabular}{@{}lcc@{}}
\toprule
& \multicolumn{2}{c}{Throughput (kS/s)} \\
\cmidrule(l){2-3}
Fitter Algorithm & Apple M4 Max\tnotex{tn:m4} & Intel Core Ultra 7\tnotex{tn:i7} \\
\midrule
\texttt{StandardNLS}, sequential & 3,804 & 1,889 \\
\texttt{StandardNLS}, parallel & 13,334 & 11,940 \\
\textit{(Speedup)} & \textit{(3.5x)} & \textit{(6.3x)} \\
\addlinespace
\texttt{StandardEKF} & 904 & 726 \\
\texttt{IntegratedEKF} & 398 & 310 \\
\bottomrule
\end{tabular}
\begin{tablenotes}
    \item[$\dagger$] \label{tn:m4} 16 cores (12P+4E), up to 4.5\,GHz. Tested on macOS 15.5.
    \item[$\ddagger$] \label{tn:i7} 16 cores (6P+8E+2LP), up to 4.8\,GHz. Tested on Ubuntu 24.04 via WSL-2 on Windows 11 Pro.
\end{tablenotes}
\end{table}

The benchmark demonstrates that the parallelized NLS fitter achieves the highest overall throughput, making it ideal for the rapid offline analysis of large datasets. The JIT-compiled EKF fitters also achieve very high throughputs, well in excess of typical data acquisition rates, confirming their suitability for real-time applications. The trade-off between model complexity and computational cost is evident: the simpler 5D random walk model offers more than double the throughput of the more complex 10D constant velocity model. This allows researchers to select the appropriate real-time algorithm based on their specific needs for tracking fidelity versus computational overhead.

\section{High-Fidelity Physics Engine}
\label{sec:physics}
A key capability of \dfmk{} is the simulation of complex interferometric signals that include dynamic path length changes and realistic, colored noise. This feature is essential for testing the tracking performance of readout algorithms and for studying the coupling of various physical effects into the DFMI signal. The implementation, located within the \texttt{SignalGenerator} class in \texttt{physics.py}, is designed to model the underlying physics with high fidelity, moving beyond the idealized model of Section~\ref{sec:theory}. This section details the simulation pipeline, with a focus on how time-varying delays and noise are handled.

\begin{algorithm}[b!]
\caption{High-fidelity physics simulation pipeline}
\label{alg:dynamic_sim}
\begin{algorithmic}[1]
\STATE \textbf{Input:} Simulation config $\mathcal{C}$, noise arrays $\mathcal{N}$, time axis $t$
\STATE \textbf{Phase 1. Calculate Total Dynamic Path and Delays}
\STATE $l_{r,0}(t) \leftarrow \mathcal{C}.\text{\texttt{IfoConfig.ref\_arml}}$
\STATE $l_{m,0}(t) \leftarrow \mathcal{C}.\text{\texttt{IfoConfig.meas\_arml}}$
\STATE $l_{\text{sig}}(t) \leftarrow \mathcal{C}.\text{\texttt{IfoConfig.}\{prescribed physical motion\}}$
\STATE $l_n(t) \leftarrow \mathcal{N}(\text{\texttt{IfoConfig.arml\_n}})$
\STATE $\text{OPD}_{\text{dyn}}(t) \leftarrow l_{\text{sig}}(t) + l_n(t)$
\STATE $\tau_r(t) \leftarrow l_{r,0}/c$; $\tau_m(t) \leftarrow (l_{m,0} + \text{OPD}_{\text{dyn}}(t))/c$
\STATE \textbf{Phase 2. Generate Base Phase Modulation Waveform}
\STATE $\Delta f(t) \leftarrow \mathcal{C}.\text{\texttt{LaserConfig.df}} + \mathcal{N}(\text{\texttt{LaserConfig.df\_n}})$
\STATE $f_{\text{mod}}(t) \leftarrow \Delta f(t) \cdot \mathcal{C}.\text{\texttt{LaserConfig.waveform\_func}}$
\STATE $\phi_{\text{mod}}(t) \leftarrow 2\pi \int f_{\text{mod}}(t) dt$ \COMMENT{Integrate to get phase}
\STATE \textbf{Phase 3. Project Modulation Phase Difference}
\STATE $\text{shifts}_r \leftarrow -\tau_r(t) \cdot f_s$; $\text{shifts}_m \leftarrow -\tau_m(t) \cdot f_s$
\STATE $\phi_{\text{mod, ref}}(t) \leftarrow \text{timeshift}(\phi_{\text{mod}}(t), \text{shifts}_r)$
\STATE $\phi_{\text{mod, meas}}(t) \leftarrow \text{timeshift}(\phi_{\text{mod}}(t), \text{shifts}_m)$
\STATE $\Delta \Phi_{\text{DFMI}}(t) \leftarrow \phi_{\text{mod, ref}}(t) - \phi_{\text{mod, meas}}(t)$
\STATE \textbf{Phase 4. Project Laser Frequency Noise}
\STATE $f_n(t) \leftarrow \mathcal{N}(\text{\texttt{LaserConfig.f\_n}})$
\STATE $f_{n, \text{ref}}(t) \leftarrow \text{timeshift}(f_n(t), \text{shifts}_r)$
\STATE $f_{n, \text{meas}}(t) \leftarrow \text{timeshift}(f_n(t), \text{shifts}_m)$
\STATE $\Delta \phi_{n,f}(t) \leftarrow 2\pi \int (f_{n, \text{ref}}(t) - f_{n, \text{meas}}(t)) dt$
\STATE \textbf{Phase 5. Assemble Final Signal}
\STATE $\Delta \Phi_{\text{carrier}}(t) \leftarrow (\omega_0/c) \cdot \text{OPD}_{\text{dyn}}(t) + \mathcal{C}.\text{\texttt{IfoConfig.phi}}$
\STATE $\Delta \Phi_{\text{total}}(t) \leftarrow \Delta \Phi_{\text{carrier}}(t) + \Delta \Phi_{\text{DFMI}}(t) + \Delta \phi_{n,f}(t)$
\STATE $A(t) \leftarrow \mathcal{C}.\text{\texttt{LaserConfig.amp}} + \mathcal{N}(\text{\texttt{LaserConfig.amp\_n}})$
\STATE $v(t) \leftarrow A(t) \cdot (1 + k \cos(\Phi_{\text{total}}(t)))$
\STATE $v(t) \leftarrow v(t) + \mathcal{N}(\text{\texttt{IfoConfig.s\_n}})$
\STATE \textbf{return} $v(t)$
\end{algorithmic}
\end{algorithm}

\subsection{Time Delay Model}
The central challenge in simulating a dynamic interferometer is modeling the time-of-flight delays correctly. As described in Section~\ref{sec:general_signal}, the interferometric signal at time $t$ is a superposition of the laser field from two different points in its history, $t-\tau_m(t)$ and $t-\tau_r(t)$. When path lengths are changing, these delays become time-dependent. Furthermore, the delay in sample units, $\tau(t) \cdot f_s$, is almost never an integer.

To address this, \dfmk{} avoids the Taylor-series approximations used in the ideal model and instead implements a direct, high-fidelity time-shifting operation. This is accomplished with the \texttt{dsp.timeshift} function, which applies a fractional delay to a time series using a high-order Lagrange interpolation filter. This approach is equivalent to convolving the signal with a finite-impulse response (FIR) filter whose coefficients are precisely calculated to represent the desired time shift. By allowing the shift value to be a time-varying array, this function accurately models the effect of dynamic path length changes on the modulation waveform, capturing higher-order effects neglected in the simplified model.

\subsection{Simulation Pipeline and Noise Injection}
The process of generating a dynamic signal is captured in the \texttt{\_run\_physics\_simulation} method. The sequence of operations, summarized in Algorithm~\ref{alg:dynamic_sim}, ensures that all physical effects and noise sources are introduced at the correct stage of the signal chain. The pipeline is designed to inject noise at physically appropriate points, correctly capturing their differential effects:
\begin{itemize}
    \item \textbf{Interferometer path length noise} (\texttt{arml\_n}, in $\rm m/\sqrt{Hz}$): This noise is treated as a physical fluctuation of the OPD. It is added directly to the prescribed path length motion before the time-of-flight delays are calculated, ensuring it correctly modulates both the carrier phase and the DFMI phase term.
    
    \item \textbf{Laser tuning amplitude noise} (\texttt{df\_n}, in $\rm Hz/\sqrt{Hz}$): This noise is applied to the modulation amplitude \textit{before} the base phase modulation waveform is generated. It therefore scales the entire $\phi_{\text{mod}}(t)$ waveform before it is time-shifted.
    
    \item \textbf{Laser carrier frequency noise} (\texttt{f\_n}, in $\rm Hz/\sqrt{Hz}$): This is a critical case. A single time series for the frequency noise is generated. It is then time-shifted independently for the reference and measurement arms, just as the modulation waveform is. The differential frequency noise is subsequently integrated to find its contribution to the final interferometric phase. This correctly models the partial cancellation of common-mode laser frequency noise in an unequal-arm interferometer.
    
    \item \textbf{Relative intensity noise} (\texttt{r\_n}, in $\rm 1/\sqrt{Hz}$): This noise, also known as RIN, is modeled as a multiplicative term affecting the overall signal amplitude, modulating both its DC offset and its AC component.
    
    \item \textbf{Sensing noise} (\texttt{s\_n}, in $\rm V/\sqrt{Hz}$): This term represents noise sources in the photo-detection process, such as shot noise or electronic noise. It is modeled as an additive noise source, applied to the signal in the final stage.
\end{itemize}

\subsection{Boundary Condition Handling}
A practical challenge of using FIR filters for time-shifting is the handling of boundary conditions. The interpolation filter requires a window of samples (e.g., 32 samples for a 31st-order filter) centered on the point being calculated. For samples near the beginning or end of a dataset, this window would extend beyond the data's boundaries.

To prevent such artifacts, \dfmk{} employs a robust ``pad-and-crop'' methodology. Instead of simulating the exact number of samples requested, the engine first calculates the required padding based on the maximum possible time delay and the FIR filter order. It then generates a longer time series that includes these extra padding cycles at both the start and end. The entire simulation pipeline, including time-shifting and noise injection, is performed on this extended, padded array. In the final step, the now-invalid padding regions are discarded, and only the central, valid portion of the simulated signal is returned to the user. This ensures that every sample in the final output has been calculated using a full, valid interpolation window, guaranteeing the integrity of the simulation across its entire duration.

\section{High-Throughput Studies: The \texttt{Experiment} Framework}
\label{sec:experiments}

To facilitate large-scale computational studies, such as multi-dimensional parameter sweeps and Monte Carlo analyses, \dfmk{} provides a high-level framework encapsulated in the \texttt{experiments.py} and \texttt{factories.py} modules. This system automates the generation, execution, and aggregation of results from thousands of individual simulation trials, leveraging parallel processing to minimize execution time.

\subsection{The Factory Pattern for Code Portability}

A significant challenge in parallel computing with Python, particularly in interactive environments like Jupyter notebooks, is the serialization (or ``pickling'') of code to be sent to worker processes. Functions and classes defined in the main script's global scope often cannot be pickled, leading to errors. To solve this and ensure robust parallel execution, \dfmk{} employs the Factory software pattern.

The user must encapsulate all logic for configuring a single simulation trial, such as defining a custom waveform or setting physics parameters, within a class that inherits from the abstract base class \texttt{ExperimentFactory}. This user-defined class implements a \texttt{\_\_call\_\_} method that takes a dictionary of trial-specific parameters and returns a dictionary of fully configured physics objects (e.g., \texttt{LaserConfig}, \texttt{IfoConfig}), as shown in Figure~\ref{fig:factory_example}. An instance of this factory class is then passed to the main experiment controller. Because the class instance and its methods are self-contained and pickleable, it can be reliably transmitted to and executed by any worker process.

\begin{figure*}[h!]
\centering
\begin{python}
from deepfmkit import physics, factories
from typing import Callable, Set

class MyDistortionExperiment(factories.ExperimentFactory): # User-defined ExperimentFactory class.
  def __init__(self, waveform_function: Callable, fm: float = 1e3, opd: float = 0.05): 
  self.waveform_func, self.fm, self.opd = waveform_function, fm, opd
  
  def __call__(self, params: dict) -> dict: # Contains the configuration logic for one trial.
    ifo = physics.IfoConfig(ref_arml = 0.1, meas_arml = 0.1 + self.opd)
    ifo.phi = params.get("phi", 0)         # Interferometric phase.
    las = physics.LaserConfig()
    las.fm = self.fm                        # Modulation frequency.
    las.psi = params.get("psi", 0)          # Modulation phase.
    las.waveform_func = self.waveform_func  # Modulation waveform.
    las.waveform_kwargs = {                 # Waveform parameters.
      "distortion_amp": params.get("distortion_amp", 0.0),
      "distortion_phase": params.get("distortion_phase", 0.0),
    }
    las.set_df_for_m(ifo, params['m_main']) # Set laser.df to achieve target effective modulation depth.
    return {'laser_config': laser, 'main_ifo_config': ifo} # Return dictionary of configured objects.
    
  def _get_expected_params_keys(self) -> Set[str]: # Declares parameters consumed by __call__.
    return {"m_main", "phi", "psi", "distortion_amp", "distortion_phase"}
\end{python}
\caption{Example of a user-defined \texttt{ExperimentFactory}. The logic for generating physics configurations is encapsulated within a pickle-safe class, enabling robust parallel execution.}
\label{fig:factory_example}
\end{figure*}

\subsection{The Declarative Experiment Controller}
The \texttt{Experiment} class is the primary user interface for this framework. It provides a declarative API for defining the structure of a computational experiment. Rather than writing procedural loops, the user specifies the experiment's components:
\begin{itemize}
    \item \texttt{add\_axis(name, values)}: Defines a parameter to be swept over the given array of \texttt{values}. Multiple axes can be added to create an N-dimensional parameter grid.
    \item \texttt{set\_static(params)}: Sets parameters that remain constant across all trials.
    \item \texttt{add\_stochastic\_variable(name, generator)}: Defines a parameter for Monte Carlo analysis, whose value is drawn from a provided \texttt{generator} function for each trial.
    \item \texttt{add\_analysis(name, fitter\_method, ...)}: Specifies a readout algorithm (e.g., \texttt{StandardNLS}) to be run on each simulated dataset.
\end{itemize}
The user then associates an instance of their custom factory with the experiment using \texttt{set\_config\_factory()}, as shown in Figure~\ref{fig:distortion-experiment}.

\subsection{Execution and Result Aggregation}
The \texttt{Experiment.run()} method orchestrates the entire simulation campaign. Its logic, summarized in Algorithm~\ref{alg:experiment_run}, is divided into three main phases: job generation, parallel execution, and result aggregation.

\begin{algorithm}
\caption{\texttt{Experiment.run()} execution pipeline}
\label{alg:experiment_run}
\begin{algorithmic}[1]
\STATE \textbf{Phase 1: Job Generation}
\STATE Initialize empty list \texttt{job\_packets}
\STATE \texttt{grid\_points} $\leftarrow$ \texttt{itertools.product}(\texttt{axes.values()})
\FOR{each \texttt{point} in \texttt{grid\_points}}
    \FOR{$i=1$ \TO \texttt{n\_trials}}
        \STATE \texttt{params} $\leftarrow$ \texttt{static\_params}
        \STATE Add \texttt{point} values to \texttt{params}
        \STATE Generate stochastic variables and add to \texttt{params}
        \STATE Create \texttt{packet} = (\texttt{params}, \texttt{factory}, \texttt{analyses})
        \STATE Append \texttt{packet} to \texttt{job\_packets}
    \ENDFOR
\ENDFOR
\STATE \textbf{Phase 2: Parallel Execution}
\STATE Create \texttt{multiprocessing.Pool} with \texttt{n\_cores}
\STATE \texttt{flat\_results} $\leftarrow$ \texttt{pool.map}(\texttt{\_run\_single\_trial}, \texttt{job\_packets})
\STATE \textbf{Phase 3: Result Aggregation}
\STATE Initialize N-dimensional result arrays with NaNs
\FOR{each \texttt{result} in \texttt{flat\_results}}
    \STATE Get grid index and trial index from \texttt{result.params}
    \STATE Place analysis values into correct location in N-D arrays
\ENDFOR
\STATE Compute statistics (mean, std) across trial dimensions of arrays, store results
\STATE \textbf{return} structured results dictionary
\end{algorithmic}
\end{algorithm}

During job generation, the controller creates the Cartesian product of all defined axes to form a grid of experimental points. For each point, it generates job ``packets'' for each Monte Carlo trial. Each packet is a self-contained tuple holding all the information needed for one atomic ``simulate-and-analyze'' run.

During execution, these packets are distributed to a pool of worker processes. Each worker receives a packet, uses the factory to configure the physics, runs the simulation, performs the requested analyses, and returns a dictionary of results.

Finally, during aggregation, the main process collects the flat list of results from all workers and reconstructs structured, N-dimensional NumPy arrays corresponding to the experiment's axes. This final data structure contains the results of all individual trials, as well as summary statistics such as the mean and standard deviation across the Monte Carlo dimension, ready for plotting and further analysis.

\section{Illustrative Application: Second Harmonic Distortion}
\label{sec:application}

\begin{figure*}[htpb!]
\centering
\begin{python}
from deepfmkit.experiments import Experiment
from deepfmkit.factories import StandardDFMIExperimentFactory
from deepfmkit.waveforms import shd
import numpy as np

# --- Declare experiment factory --- 
fcty = StandardDFMIExperimentFactory(
  waveform_function=shd, # Use the 'second harmonic distortion' modulation waveform
  fm  = 1e3              # Set the laser's modulation frequency to 1 kHz
  opd = 0.05,            # Set the interferometer's OPD to 5 cm
)

# --- Define experiment parameters --- 
exp = Experiment(description="2nd Harmonic Distortion")
exp.set_config_factory(fcty)      # Set desired ExperimentFactory
exp.f_samp = 200e3                # Sampling frequency
exp.n_cycles = 10                 # Modulation cycles per fit buffer
exp.n_trials = 500                # Number of Monte Carlo trials

# --- Parameter sweeps --- 
exp.add_axis("m_main", np.linspace(3.0, 30.0, 1000))        # 1. Effective modulation depth (rad)
exp.add_axis("distortion_amp", np.linspace(0.0, 0.2, 100))  # 2. Harmonic distortion amplitude (fractional)

# --- Variables for Monte Carlo experiment --- 
exp.add_stochastic_variable("distortion_phase", lambda: np.random.uniform(-np.pi, np.pi))
exp.add_stochastic_variable("phi", lambda: np.random.uniform(-np.pi, np.pi))

# --- Analysis to perform for each trial --- 
exp.add_analysis(
    name="My Analysis",
    fitter_method="nls",               # Run the StandardNLS
    result_cols=["m", "phi"],          # Collect the outputs `m` and `phi`
    fitter_kwargs={"n_harmonics": 30}  # Use a 30-harmonic model
)

# --- Run experiment --- 
exp.results = exp.run(filename="experiment.data")
\end{python}
\caption{Example high-throughout \texttt{Experiment} script that makes use of the \texttt{StandardDFMIExperimentFactory} included in \texttt{factories.py}. The logic for generating physics configurations is encapsulated within a pickle-safe class, enabling robust parallel execution of the 50,000,000 trial runs in this example.}
\label{fig:distortion-experiment}
\end{figure*}

\begin{figure*}[htpb!]
\centering
\includegraphics{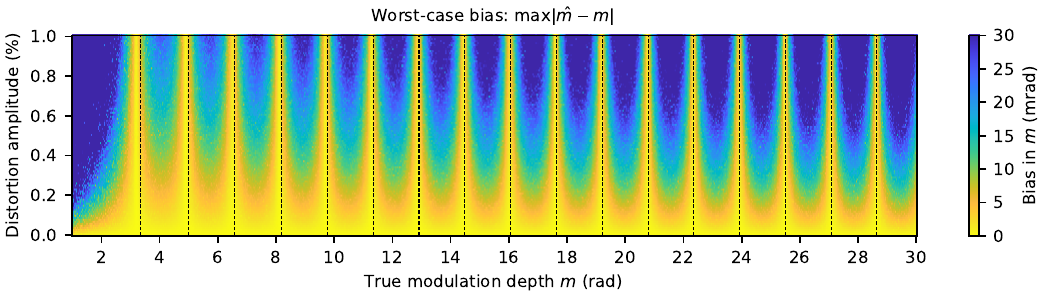}
\caption{Results of the high-throughput experiment from Figure~\ref{fig:distortion-experiment}, showing the computed worst-case bias in the estimated modulation depth ($\hat{m}$) due to second-harmonic distortion in the frequency modulation. The bias is shown as a function of the true modulation depth $m$ and the distortion amplitude $\epsilon$. The worst-case bias results from a Monte Carlo simulation over the unknown phases $\Phi$ and $\psi_2$. The plot reveals distinct vertical ``valleys of robustness'' where the bias is strongly suppressed, which are predicted analytically to align with the extrema of $J_2(2m)$ starting at $m=3.35$ (i.e., occurring at $m \approx 3.35, 4.98, 6.59, 8.17 \dots$, indicated by dashed black lines).}
\label{fig:distortion-bias}
\end{figure*}

To demonstrate the utility of the \dfmk{} framework for systematic characterization, we present a high-throughput study investigating the robustness of the NLS readout algorithm to second-harmonic distortion (SHD) in the laser's modulation waveform. 

Systematic errors can arise from imperfections in laser frequency modulation. A frequency modulator may exhibit harmonic distortion, deviating from the ideal sinusoidal model of Eq.~\eqref{eq:f_mod} and introducing bias into parameter estimates. The goal of this virtual experiment is to quantify the bias introduced in the estimation of the modulation depth, $\hat m$, as a function of the distortion amplitude.

We consider a distorted instantaneous frequency modulation containing a second-harmonic term, a common manifestation of nonlinearity:
\begin{equation}
f_{\rm mod}(t) = \Delta f \left[ \cos(\omega_m t + \psi) + \epsilon \cos(2\omega_m t + \psi_2) \right],
\label{eq:f_mod_distorted}
\end{equation}
where $\epsilon \ll 1$ is the relative distortion amplitude and $\psi_2$ is its phase. The total phase modulation is the time integral of the angular frequency, $2\pi f_{\rm mod}(t)$. Using a first-order Taylor approximation, the differential interferometric phase is $\Delta\phi(t) \approx 2\pi\tau f_{\rm mod}(t)$, which contains the parasitic term:
\begin{equation}
\delta\phi(t) = m \epsilon \cos(2\omega_m t + \psi_2).
\end{equation}
For a small perturbation ($\epsilon \ll 1$), the resulting error in the measured voltage, $\delta v(t)$, is approximated by:
\begin{equation}
\delta v(t) \approx -C m \epsilon \cos(2\omega_m t + \psi_2) \sin(\Phi + m\cos(\omega_m t + \psi)).
\label{eq:nl_perturbation_signal}
\end{equation}
This expression shows that the error originates from frequency mixing between the $2\omega_m$ distortion and the harmonic structure of the ideal DFMI signal. In the frequency domain, this mixing appears as spectral leakage, where power from harmonics at $(n \pm 2)f_m$ contaminates those at $nf_m$. The resulting bias in parameter estimates depends on this detailed spectral perturbation.

The experiment was configured using the \texttt{Experiment} class, as shown in the script in Figure~\ref{fig:distortion-experiment}. We defined a 2D parameter sweep over the true effective modulation depth, $m$, and the distortion amplitude, $\epsilon$. For each point on this grid, 500 Monte Carlo trials were performed. In each trial, the interferometric phase $\Phi$ and the distortion phase $\psi_2$ were drawn from a uniform random distribution between $-\pi$ and $\pi$ to average over all possible phase relationships.

An analytical derivation of the resulting bias is complex, as it depends on the precise interplay between the distortion's amplitude and phase and the main signal parameters. However, the \texttt{Experiment} framework allows this effect to be characterized numerically with high precision. The results, shown in Figure~\ref{fig:distortion-bias}, reveal distinct ``robustness valleys'' where the estimation bias is strongly suppressed.

These valleys arise from orthogonality conditions in the time domain. A parameter bias is minimized when the perturbation $\delta v(t)$ is orthogonal to the time-domain gradient of the ideal signal with respect to that parameter, averaged over a modulation period $T$. For the modulation depth $m$, this gradient is:
\begin{align}
g_m(t) &= \partial v/\partial m \nonumber \\
&= -C \cos(\omega_m t + \psi) \sin(\Phi + m\cos(\omega_m t + \psi)).
\end{align}
The zero-bias condition,
\begin{equation}
\langle \delta v(t), g_m(t) \rangle = (1/T)\int_0^T \delta v(t) g_m(t) \, dt = 0,
\end{equation}
reduces, after averaging over all unknown phases, to the requirement that the derivative of the second-order Bessel function vanishes: $J'_2(2m) = 0$~\cite{Dovale2025}. This condition predicts valleys at $m$ values corresponding to the extrema of $J_2(2m)$, matching the simulation results and correctly giving the observed asymptotic spacing of $\pi/2$.

This type of quantitative analysis is crucial for setting design specifications for the purity of modulation electronics and for understanding the performance limits of a given readout algorithm. The entire experiment, comprising 50 million unique ``simulate-and-fit'' trials, was automated by the framework and executed in parallel, demonstrating its power for the efficient and systematic investigation of complex error sources.

\section{Verification and Availability}
\label{sec:verification}

Ensuring the correctness and reliability of a scientific software library is paramount. The \dfmk{} framework has been rigorously validated through a comprehensive suite of automated tests.

\paragraph{Software Testing} The codebase is accompanied by an extensive test suite developed using the \texttt{pytest} framework, comprising over 50 unit, integration, and regression tests. Unit tests validate individual components in isolation, such as the mathematical correctness of the Lagrange interpolation in \texttt{dsp.py} and the statistical properties of the noise generators in \texttt{noise.py}. Integration tests verify that modules work together as expected; for example, a key test confirms that the NLS fitter can recover ground-truth parameters from a noiseless, ideal signal generated by the physics engine to a relative precision of $10^{-6}$. Finally, a ``golden file'' regression test runs a complete, complex experiment and compares its numerical output to a version-controlled reference file, ensuring that future code changes do not inadvertently alter scientific results. This multi-layered testing strategy provides a strong guarantee of the software's correctness.

\paragraph{Code availability} \dfmk{} is open-source software and its source code, example scripts, documentation, and the complete test suite are publicly available on GitHub at \href{https://github.com/mdovale/DeepFMKit}{https://github.com/mdovale/DeepFMKit}. The library is distributed under the permissive BSD 3-Clause license, allowing for its use in both academic and commercial applications. The software can be installed directly from the Python Package Index (PyPI) at \href{https://pypi.org/project/deepfmkit/}{https://pypi.org/project/deepfmkit/}.

\section{Summary and Outlook}
\label{sec:summary-and-outlook}

We have presented \dfmk{}, an open-source Python library for the end-to-end simulation and analysis of Deep Frequency Modulation Interferometry systems. Motivated by the need for a comprehensive and flexible modeling tool in precision metrology, we developed a framework that integrates a high-fidelity physics engine with a suite of robust parameter estimation algorithms.

The key contributions of \dfmk{} are its modular design and its powerful, validated components. The physics engine accurately models the complex signal dynamics of DFMI, including time-of-flight effects, user-defined modulation waveforms, and colored noise from $1/f^\alpha$ power spectra. For parameter estimation, the library provides two distinct readout strategies: a high-throughput, parallelized Non-linear Least Squares fitter for precise offline analysis, and real-time Extended Kalman Filters for tracking dynamic systems. This is complemented by a high-level \texttt{Experiment} framework that enables the automation of large-scale computational studies for systematic characterization. We demonstrated the utility of this framework by investigating the impact of modulation waveform distortion on parameter estimation bias. The entire library is supported by a comprehensive test suite that ensures its correctness and reliability.

Looking forward, \dfmk{} provides a powerful platform for future research. Its modular architecture readily allows for the extension of its capabilities. Future work could include integrating more sophisticated noise models, such as stray light, or expanding the fitter suite with other advanced algorithms like particle filters or machine learning-based approaches. The high-throughput experimentation framework can then be leveraged to perform systematic comparisons of these new algorithms against the established NLS and EKF benchmarks.

By providing the community with a well-documented, rigorously tested, and extensible open-source tool, we hope that \dfmk{} will serve as a valuable resource for both education and research, lowering the barrier to entry for studying DFMI and accelerating innovation in high-precision laser interferometry.

\section*{Acknowledgements}

The author gratefully acknowledges Gerhard Heinzel for his pioneering work in Deep Phase Modulation Interferometry, and in particular for developing the first specialized Levenberg-Marquardt Algorithm program, which forms the basis of the \texttt{StandardNLS} implementation.

\section{Configuration Parameters}
\label{sec:params}

Table~\ref{tab:params} provides a reference for the key configurable parameters in the \dfmk{} physics and parameter estimation engine, their physical meaning, and their units or type.

\begin{table*}[h!]
\centering
\caption{Configuration parameters for the physics and parameter estimation engines.}
\label{tab:software_params}
\begin{tabularx}{\textwidth}{@{} l X r @{}}
\toprule
\textbf{Parameter} & \textbf{Description} & \textbf{Units} \\
\midrule
\multicolumn{3}{@{}l}{\textbf{Laser Parameters (\texttt{LaserConfig})}} \\
\texttt{wavelength} & Laser carrier wavelength, which determines the optical frequency $f_0$. & nm \\
\texttt{fm} & The frequency of the laser's deep frequency modulation, $f_m$. & Hz \\
\texttt{df} & The tuning amplitude of the laser's frequency modulation, $\Delta f$. & Hz \\
\texttt{psi} & A static phase offset for the modulation waveform, $\psi$. & rad \\
\texttt{amp} & A term proportional to laser power, affecting the signal's DC offset $A$ and AC amplitude $C$. & V \\
\texttt{waveform\_func} & A Python callable that defines the shape of the modulation waveform. & (function) \\
\addlinespace
\multicolumn{3}{@{}l}{\textbf{Interferometer Parameters (\texttt{IfoConfig})}} \\
\texttt{ref\_arml} & Static length of the reference arm. & m \\
\texttt{meas\_arml} & Static length of the measurement arm. The static OPD is $\Delta l = \text{meas\_arml} - \text{ref\_arml}$. & m \\
\texttt{visibility} & Interferometer fringe visibility or contrast, $k$. & (unitless) \\
\texttt{phi} & A static offset added to the total interferometric phase ($\Phi$). & rad \\
\texttt{arml\_mod\_amp} & Amplitude of a prescribed sinusoidal motion applied to the OPD. & m \\
\texttt{arml\_mod\_f} & Frequency of the prescribed sinusoidal OPD motion. & Hz \\
\addlinespace
\multicolumn{3}{@{}l}{\textbf{Noise Parameters (Amplitude Spectral Density (ASD) @ 1 Hz and Frequency Scalings ($\text{PSD} = \text{ASD}^2 \propto \frac{1}{f^{\alpha}}$))}} \\
\textit{Laser Noises} \\
\texttt{f\_n} & Laser carrier frequency noise. & Hz/\(\sqrt{\text{Hz}}\) \\
\texttt{f\_n\_alpha} & Frequency scaling exponent ($\alpha$) for laser frequency noise. & (unitless) \\
\texttt{df\_n} & Laser tuning amplitude noise. & Hz/\(\sqrt{\text{Hz}}\) \\
\texttt{df\_n\_alpha} & Frequency scaling exponent ($\alpha$) for tuning amplitude noise. & (unitless) \\
\texttt{r\_n} & Relative intensity noise (RIN). & 1/\(\sqrt{\text{Hz}}\) \\
\texttt{r\_n\_alpha} & Frequency scaling exponent ($\alpha$) for relative intensity noise. & (unitless) \\
\addlinespace
\textit{Interferometer Noises} \\
\texttt{arml\_n} & OPD noise, representing physical fluctuations in the path length difference. & m/\(\sqrt{\text{Hz}}\) \\
\texttt{arml\_n\_alpha} & Frequency scaling exponent ($\alpha$) for OPD noise. & (unitless) \\
\texttt{s\_n} & Sensing noise, representing additive noise sources like electronic or shot noise. & V/\(\sqrt{\text{Hz}}\) \\
\texttt{s\_n\_alpha} & Frequency scaling exponent ($\alpha$) for sensing noise. & (unitless) \\
\addlinespace
\multicolumn{3}{@{}l}{\textbf{Simulation Parameters (\texttt{SimConfig})}} \\
\texttt{f\_samp} & The sampling frequency of the simulation and data acquisition. & Hz \\
\addlinespace
\multicolumn{3}{@{}l}{\textbf{Parameter Estimation}} \\
\multicolumn{3}{@{}l}{Non-linear Least Squares (\texttt{StandardNLS})} \\
\texttt{n\_cycles} & Number of modulation cycles of raw data per fit buffer. & (integer) \\
\texttt{n\_harmonics} & Number of harmonics $\alpha_n$ to use in the NLS signal model. & (integer) \\
\texttt{fit\_params} & A list specifying which parameters (\texttt{'amp'}, \texttt{'m'}, \texttt{'phi'}, \texttt{'psi'}) to include in the fit. & (list) \\
\texttt{init\_psi\_method} & Method for automated initialization of the modulation phase, $\psi$. & (string) \\
\texttt{fs} & Dara rate of the output. Determined by \texttt{n\_cycles}, \texttt{f\_samp}, and \texttt{fm}. & (float) \\
\addlinespace
\multicolumn{3}{@{}l}{Extended Kalman Filters (\texttt{StandardEKF}, \texttt{IntegratedEKF})} \\
\texttt{P0\_diag} & Diagonal elements of the initial state error covariance matrix, $\mathbf{P}_{0|0}$. & (various) \\
\texttt{Q\_diag} & Diagonal elements of the process noise covariance matrix, $\mathbf{Q}$. Defines the filter's dynamics. & (various) \\
\texttt{R\_val} & The measurement noise variance, $R$. & V\(^2\) \\
\bottomrule
\end{tabularx}
\label{tab:params}
\end{table*}

\section{Class Diagrams}
\label{sec:class-diagrams}

This appendix provides a visual overview of the software architecture through UML class diagrams. The diagrams illustrate the main structural components of the \dfmk{} library and their interactions, highlighting the separation of responsibilities and modular design that facilitate extensibility.

\begin{figure*}[t!]
    \centering
    \begin{tikzpicture}[
        node distance=1cm,
        class/.style={rectangle, rounded corners, draw=black, thick, fill=blue!10, minimum height=2.5cm, text width=4cm, align=left, drop shadow},
        data/.style={class, fill=orange!10, text width=3.5cm},
        arrow/.style={->, thick}
    ]
    \node (dff) [class, text width=5cm] {
        \textbf{DeepFrame} \\
        \hrulefill \\
        \texttt{- sims: dict[SimConfig]} \\
        \texttt{- raws: dict[RawData]} \\
        \texttt{- fits: dict[FitData]} \\
        \texttt{- fits\_df: dict[DataFrame]} \\
        \hrulefill \\
        \texttt{+ add\_sim(sim)} \\
        \texttt{+ simulate(label, n\_seconds)} \\
        \texttt{+ fit(main\_label, method)} \\
        \texttt{+ plot(labels)} \\
        \texttt{+ load\_raw(filepath)} \\
        \texttt{+ load\_fit(filepath)}
    };
    \node (sim) [class, fill=green!10, right=1.5cm of dff, text width=4cm] {
        \textbf{SimConfig} \\
        \hrulefill \\
        \texttt{- label: str} \\
        \texttt{- laser: LaserConfig} \\
        \texttt{- ifo: IfoConfig} \\
        \texttt{- f\_samp: float} \\
        \hrulefill \\
        \texttt{+ m: float (property)} \\
        \texttt{+ info()} \\
        \texttt{+ plot\_harmonics()}
    };
    \node (raw) [data, right=2.5cm of sim] {
        \textbf{RawData} \\
        \hrulefill \\
        \texttt{- label: str} \\
        \texttt{- data: DataFrame} \\
        \texttt{- f\_samp: float} \\
        \texttt{- fm: float} \\
        \texttt{- sim: SimConfig} \\
        \hrulefill \\
        \texttt{+ to\_txt(filename)} \\
        \texttt{+ plot()}
    };
    \node (fit) [data, below=of raw] {
        \textbf{FitData} \\
        \hrulefill \\
        \texttt{- label: str} \\
        \texttt{- amp: array} \\
        \texttt{- m: array} \\
        \texttt{- phi: array} \\
        \texttt{- psi: array} \\
        \texttt{- fs: float} \\
        \hrulefill \\
        \texttt{+ to\_txt(filename)} \\
        \texttt{+ plot()}
    };
    \draw[arrow] (dff.east) -- ++(1.25,0) node[midway, above] {\texttt{uses}} |- (sim.west);
    \draw[arrow, dashed] (sim) -- node[midway, above] {physics engine} (raw);
    \draw[arrow, dashed] (raw) -- node[midway, right] {fitter} (fit);
    \end{tikzpicture}
    \caption{Core workflow diagram for configuring and running physics simulations. The central \texttt{DeepFrame} class orchestrates the workflow, using a \texttt{SimConfig} object to generate \texttt{RawData}, which is then processed by a fitter to produce \texttt{FitData}.}
    \label{fig:diag_core}
\end{figure*}
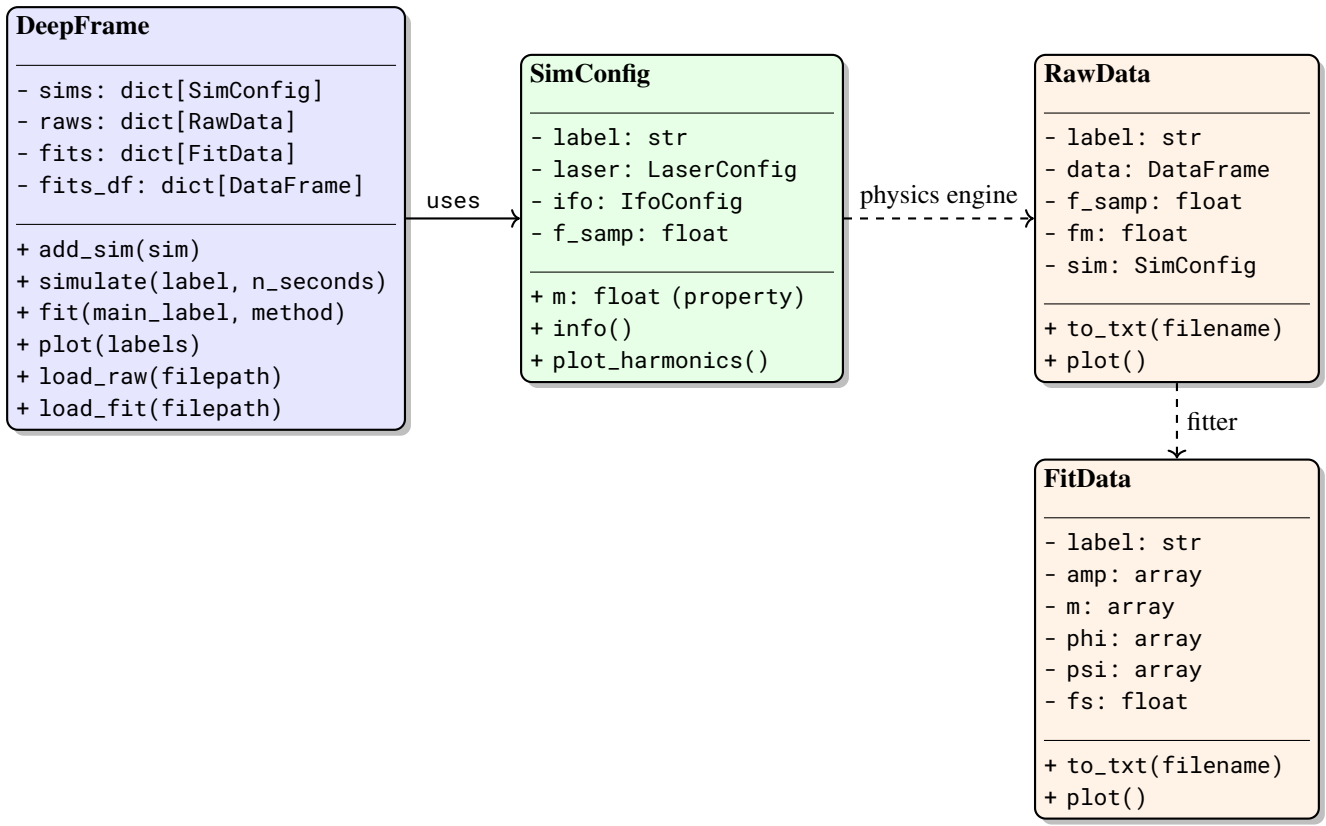

Figure~\ref{fig:diag_core} presents the high-level library workflow. The central controller, \texttt{DeepFrame}, manages the execution pipeline by coordinating physics simulations, data handling, and parameter estimation. It uses configuration objects (\texttt{SimConfig}) to define simulation parameters and fitting strategy classes (derived from \texttt{BaseFitter}) to perform analysis.

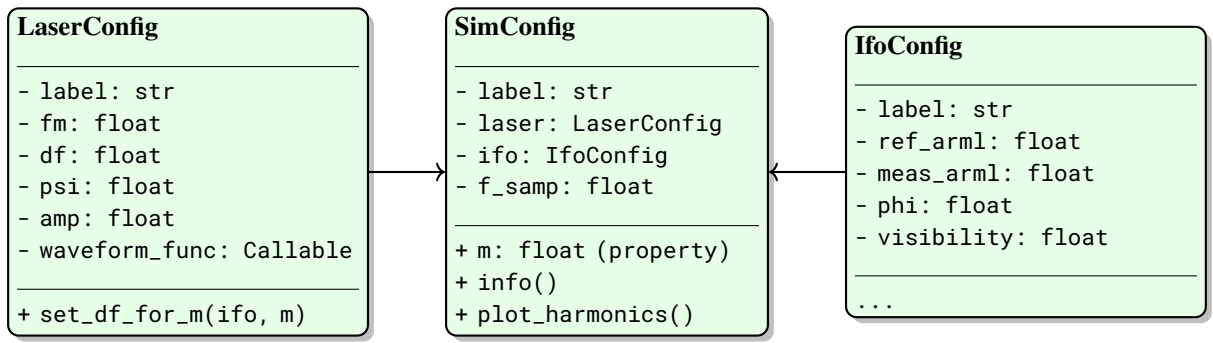
\begin{figure*}[t!]
    \centering
    \begin{tikzpicture}[
        node distance=1cm,
        class/.style={rectangle, rounded corners, draw=black, thick, fill=green!10, minimum height=2.5cm, text width=4.5cm, align=left, drop shadow},
        compo/.style={draw, thick, ->}
    ]
    \node (sim) [class, text width=4cm] {
        \textbf{SimConfig} \\
        \hrulefill \\
        \texttt{- label: str} \\
        \texttt{- laser: LaserConfig} \\
        \texttt{- ifo: IfoConfig} \\
        \texttt{- f\_samp: float} \\
        \hrulefill \\
        \texttt{+ m: float (property)} \\
        \texttt{+ info()} \\
        \texttt{+ plot\_harmonics()}
    };
    \node (laser) [class, left=of sim] {
        \textbf{LaserConfig} \\
        \hrulefill \\
        \texttt{- label: str} \\
        \texttt{- fm: float} \\
        \texttt{- df: float} \\
        \texttt{- psi: float} \\
        \texttt{- amp: float} \\
        \texttt{- waveform\_func: Callable} \\
        \hrulefill \\
        \texttt{+ set\_df\_for\_m(ifo, m)}
    };
    \node (ifo) [class, right=of sim] {
        \textbf{IfoConfig} \\
        \hrulefill \\
        \texttt{- label: str} \\
        \texttt{- ref\_arml: float} \\
        \texttt{- meas\_arml: float} \\
        \texttt{- phi: float} \\
        \texttt{- visibility: float} \\
        \hrulefill \\
        \texttt{...}
    };
    \draw[compo] (laser.east) -- (sim.west);
    \draw[compo] (ifo.west) -- (sim.east);
    \end{tikzpicture}
    \caption{Physics model composition. A \texttt{SimConfig} object is composed of one \texttt{LaserConfig} and one \texttt{IfoConfig}. The user configures the physical properties (like modulation frequency \texttt{fm} or arm length \texttt{ref\_arml}) on the component objects.}
    \label{fig:diag_physics}
\end{figure*}

Figure~\ref{fig:diag_physics} illustrates the composition of the physics model. A \texttt{SimConfig} object encapsulates an experimental configuration by composing a \texttt{LaserConfig} and an \texttt{IfoConfig} object. This design enables the independent modification and reuse of laser and interferometer models.

\begin{figure*}[t!]
    \centering
    \begin{tikzpicture}[
        node distance=1cm,
        class/.style={rectangle, rounded corners, draw=black, thick, fill=purple!10, minimum height=1.5cm, text width=4.0cm, align=left, drop shadow},
        abstract/.style={class, font=\itshape},
        inherit/.style={draw, thick, ->}
    ]
    \node (base) [abstract] {
        \textbf{BaseFitter} \\
        \hrulefill \\
        \texttt{- config: dict} \\
        \hrulefill \\
        \texttt{+ fit(main\_raw, ...)}
    };
    \node (nls) [class, below left=of base, xshift=2cm, text width=8cm] {
        \textbf{StandardNLS} \\
        \hrulefill \\
        \texttt{- ndata: int (number of harmonics in the model)} \\
        \texttt{- fit\_params: array (parameters to fit)} \\
        \texttt{- init\_psi\_method: str (smart initialization method for the modulation phase)} \\
        \hrulefill \\
        \texttt{+ fit(main\_raw, ndata, parallel, ...)}
    };
    \node (ekf) [class, below right=of base, xshift=-2cm, text width=8cm] {
        \textbf{StandardEKF / IntegratedEKF} \\
        \hrulefill \\
        \texttt{- P0\_diag: array (initial state covariance)} \\
        \texttt{- Q\_diag: array (proc.\ noise covariance)} \\
        \texttt{- R\_val: float (measurement noise variance)} \\ 
        \hrulefill \\
        \texttt{+ fit(main\_raw, P0\_diag, Q\_diag, R\_val, ...)}
    };
    \draw[inherit] (nls.north) -- (base.south west);
    \draw[inherit] (ekf.north) -- (base.south east);
    \end{tikzpicture}
    \caption{The Fitter Strategy Pattern. Concrete fitters like \texttt{StandardNLS} and \texttt{StandardEKF} inherit from a common \texttt{BaseFitter} interface, allowing them to be used interchangeably by the main controller.}
    \label{fig:diag_fitters}
\end{figure*}
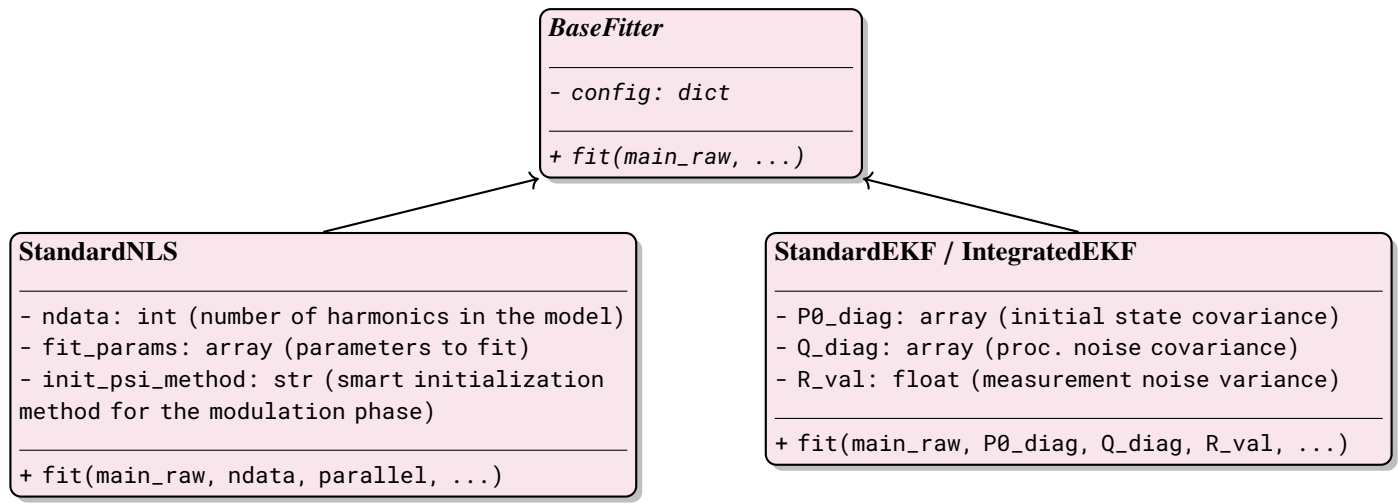

Figure~\ref{fig:diag_fitters} shows the Strategy pattern used for parameter estimation. An abstract \texttt{BaseFitter} class defines a common interface. Concrete implementations such as \texttt{StandardNLS} and \texttt{StandardEKF} provide distinct algorithms, allowing them to be used interchangeably by the main controller.

\end{document}